\documentclass{aa}  
\usepackage{graphicx}
\usepackage{lipsum}
\usepackage{txfonts}
\usepackage{comment}
\usepackage{subfigure}
\usepackage{xcolor}
\usepackage{afterpage}
\usepackage{siunitx}
\usepackage{academicons}

\begin{document} 

   \title{Tracking solar radio bursts using Bayesian multilateration}

   \author{L.~A.~Ca\~nizares \orcid{0000-0003-4711-522X}
          \inst{1,2} 
          \and
          S.~T.~Badman\inst{3} \orcid{0000-0002-6145-436X}
          \and
          S.~A.~Maloney\inst{1} \orcid{0000-0002-4715-1805}
          \and
          M.~J.~Owens\inst{4} \orcid{0000-0003-2061-2453}
          \and
          D.~M.~Weigt\inst{1,5} \orcid{0000-0001-5427-6537}
          \and
          E.~P.~Carley\inst{1} \orcid{0000-0002-6106-5292}
          \and
          P.~T.~Gallagher\inst{1} \orcid{0000-0001-9745-0400}
          }

   \institute{Astronomy \& Astrophysics Section, DIAS Dunsink Observatory, Dublin Institute for Advanced Studies, Dublin D15XR2R, Ireland.
         \and
             School of Physics, Trinity College Dublin, Dublin 2, Ireland.
         \and 
             Center for Astrophysics, Harvard \& Smithsonian, Cambridge, Massachusetts, US.
         \and 
            Department of Meteorology, University of Reading, Earley Gate, PO Box 243, Reading, RG6 6BB, UK.
        \and 
            Department of Computer Science, Aalto University, 00076 Aalto, Finland.
             }

   \date{Received August 17, 2023; accepted January 26, 2024}


  \abstract
   {Solar radio bursts (SRBs), such as Type IIs and IIIs, are emitted by electrons propagating through the corona and interplanetary space. Tracking such bursts is key to understanding the properties of accelerated electrons and radio wave propagation as well as the local plasma environment that they propagate through.}
   {In this work, we present a novel multilateration algorithm called BayEsian LocaLisation Algorithm (BELLA) and validate the algorithm using simulated and observed SRBs. In addition, apparent SRB positions from BELLA are compared with comparable localisation methods and the predictions of solar wind models.} 
   {BELLA uses Bayesian inference to create probabilistic distributions of source positions and their uncertainties. This facilitates the estimation of algorithmic, instrumental, and physical uncertainties in a quantitative manner.}
   {We validated BELLA using simulations and a Type III SRB observed by STEREO A and STEREO B at $\pm$~116~$\degr$ from the Sun-Earth line and by Wind at L1. BELLA tracked the Type III source from $\sim$~10--150~$R_\sun$ (2--0.15~MHz) along a spiral trajectory. This allowed for an estimate of an apparent solar wind speed of $v_{sw}~\sim$~400~km~s$^{-1}$ and a source longitude of $\phi_0~\sim$~30$\degr$. We compared these results with well-established methods of positioning: Goniopolarimetric (GP), analytical time-difference-of-arrival (TDOA), and Solar radio burst Electron Motion Tracker (SEMP). We found them to be in agreement with the results obtained by BELLA. Additionally, the results aligned with solar wind properties assimilated by the Heliospheric Upwind Extrapolation with time dependence (HUXt) model.} 
   {We have validated BELLA and used it to identify apparent source positions as well as velocities and densities of the solar wind. Furthermore, we identified higher than expected electron densities, suggesting that the true emission sources were at lower altitudes than those identified by BELLA, an effect that may be due to appreciable scattering of electromagnetic waves by electrons in interplanetary space.}


    \maketitle
    %
    
\nolinenumbers
\section{Introduction}

Solar radio bursts (SRBs) have been used to track energetic particles from the Sun for decades \citep[e.g.][]{tri:reiner1998winduly,tri:cecconi2007influence, tri:jebaraj2020using}. The most abundant of these SRBs are Type IIIs \citep{sw:reid2014review}, which are generated by mildly relativistic beams of electrons travelling along open magnetic field lines that then interact with the local plasma and generate Langmuir waves \citep{sw:lorfing2023solar}. These electrostatic waves decay to produce radio wave radiation at the plasma frequency or its second harmonic \citep{sw:ginzburg1958possible, sw:reid2014review}. Type III radio bursts have proven to be a productive instrument to study the properties of the solar corona \citep{sw:reid2014review} and the physical mechanisms that drive solar energetic particles (SEP) into the heliosphere \citep[e.g.][]{tri:reiner1998winduly}. Interferometric telescopes such as the Low-Frequency Array \citep[LOFAR;][]{inst:vanHaarlem2013lofar}{}{}, which operates at 10--250~MHz, allow for relatively accurate tracking of radio sources at $<3R_\odot$. For example, \cite{sw:morosan2014lofar} tracked a Type III SRB at 30--90~MHz using LOFAR tied-array observations with high temporal ($\sim$~$30$~ms) and spectral (12.5~kHz) resolution. Furthermore, LOFAR facilitates interferometric imaging in the manner carried out by \cite{sw:maguire2021lofar}, where a Type II radio source is imaged to a height of $\sim$~$0.5~R_\odot$. However, due to the Earth's atmospheric cutoff, lower frequency emission from higher altitudes cannot be observed using ground-based telescopes, and thus space-based antennas are required at these frequencies ($\leq10$~MHz).

Observations at less than $10$~MHz are particularly well suited to studying radio sources at greater distances from the Sun. With multiple spacecraft, source positions can  be triangulated using a variety of methods. A well-established method of triangulation is the goniopolarimetric (GP) method, or direction finding, \citep{tri:manning1980new} in which the Poynting $\vec{k}$ vectors of the radio waves observed by different spacecraft are back propagated until they intersect \citep[e.g.][]{tri:reiner1998winduly}. An example of this method is described in \cite{tri:magdalenic2014tracking}, who used observations from the Waves instrument on board the Wind spacecraft \citep[Wind/Waves;][]{inst:Bougeret1995waves} and the  Waves instruments on board the Solar Terrestrial Relations Observatory \citep[STEREO/Waves;][]{inst:Bougeret2008swaves} to triangulate a Type II associated with a coronal mass ejection (CME). \cite{tri:weber1977interplanetary, Steinberg1984} showed earlier time delay measurements with two spacecraft as an alternative to the GP method. With the addition of new spacecraft, the time-difference-of-arrival (TDOA) method was developed \citep{Alcock2018} for the purpose of tracking radio sources as shown in \cite{tri:badman2022tracking}, where a Type III radio burst was successfully tracked from 0.1--16~MHz using the Parker Solar Probe \citep[PSP/FIELDS;][]{inst:Bale2016Fields}, STEREO A/Waves, and Wind/Waves. An alternative method of positioning using arrival times is the Solar radio burst Electron Motion Tracker \citep[SEMP;][]{tri:zhang2019forward}, which fits a Type III to a Parker spiral using a forward modelling method. However, SEMP is constrained by the use of density models and an assumed constant solar wind speed and field line. Another innovative method of positioning is shown in \cite{tri:reiner2009multipoint} and \cite{tri:musset2021simulations}, where they used the intensity of the Type III radio sources in order to calculate the directivity of the radio emissions and density models to obtain the radial distance from the Sun. Lastly, \cite{tri:chen2023source} have combined simulations of density fluctuations with several of these methods to account for scattering of the radio waves as they escape their source.\\

These localisation methods have various advantages and disadvantages as a result of their assumptions. For example, GP has the advantage that it requires a minimum of two spacecraft in order to obtain a solution as opposed to TDOA, which requires three spacecraft. However, GP is limited, as it requires extensive knowledge of antenna geometry, spacecraft potential, and spacecraft attitude state; for example, while Wind is spinning, STEREO is three-axis stabilised, requiring different techniques to obtain the Poynting vectors \citep{tri:manning1980new, tri:krupar2012goniopolarimetric}. It also has the disadvantage that when a pair of spacecraft is at a large angle of separation (e.g. $\sim 180\degr$), the Poynting vectors might point at each other, and the method therefore does not yield a solution. An example of this case was noted by \cite{tri:jebaraj2020using}, where the STEREO A/Waves--STEREO B/Waves pair was not used due to the unreliability of the results as a consequence of the large angular separation of the spacecraft. In contrast, multilateration works best at large separation angles (see Sect.~\ref{sec:bayessimulation}). \cite{tri:badman2022tracking} showed a longitudinal displacement close to the Sun on the order of the spread in positions from the time resolution error between the results obtained by their LOFAR interferometric results and the multilateration performed using PSP, STEREO A, and Wind. This discrepancy may likely be attributable to a number of factors, physical or systematic. The free-streaming assumption is intrinsic to the TDOA method via the assumption of light travelling at the speed of light (\textit{c}), and as a consequence, the most relevant source of physical error could be radio scattering \citep{tri:chen2023source, Kontar2019}. Other sources of errors could be the low temporal resolution of the spacecraft, in particular the Wind/Waves instrument, which provides data at a 1-min cadence; the close proximity of PSP to the Sun at perihelion ($\sim$10~$R_\odot$), or a poor spacecraft configuration (discussed in Sect.~\ref{sec:bayessimulation}). \cite{tri:badman2022tracking} estimated errors via repeating the TDOA process a specific number of times with different timestamps selected from an uncertainty range informed by instrument resolution and captured only one of the error sources listed above. A more elegant solution is the application of Bayes' theorem for the purpose of applying a statistical method that would allow for a quantitative analysis of the multiple error sources.\\

In this paper, we present the BayEsian LocaLisation Algorithm, \cite[BELLA;][]{soft:luis_alberto_canizares_alberto_2023_10276815}\footnote{\url{https://github.com/TCDSolar/BELLA}} a novel method of tracking the apparent positions of SRB sources. In Sect.~\ref{sec:method}, we introduce Bayesian multilateration and describe its implementation. In Sect.~\ref{sec:bayessimulation}, we perform a simulation for the purposes of validation and choosing a suitable candidate for a use case. In Sect.~\ref{sec:observation}, we characterise a Type III radio burst, and finally in Sect.~\ref{sec:resultsanddiscussion}, we compare the results obtained by BELLA with other well-established positioning methods and a solar wind model. We also discuss some of the potential applications of BELLA for the estimation of solar wind velocities and electron densities along the Parker spiral. 

\begin{figure}
  \centering
  \includegraphics[width=0.4\textwidth]{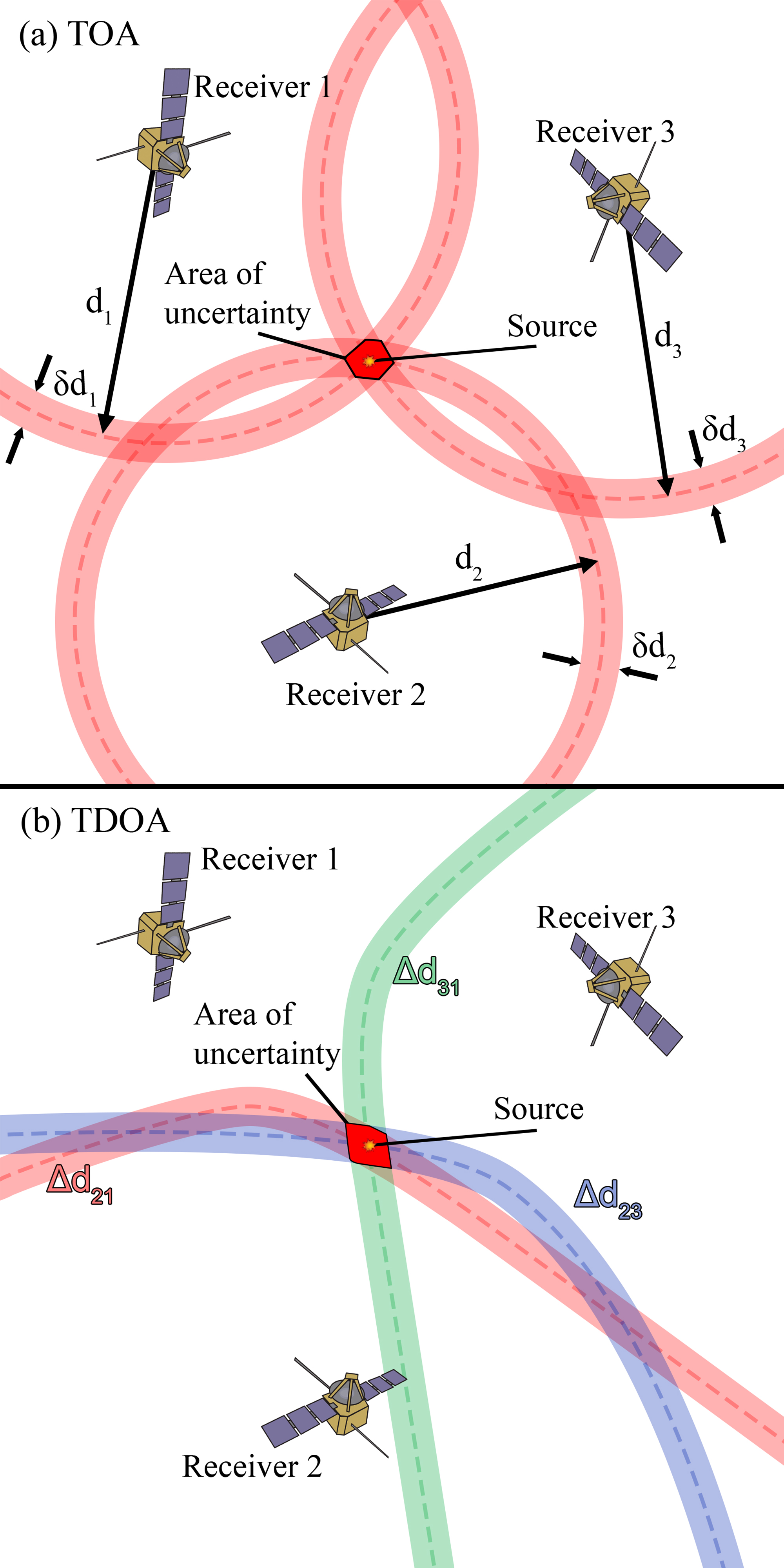}
  \caption{Time-of-arrival (TOA) method and time-difference-of-arrival (TDOA) method. (a) The TOA method uses the time difference between the time of emission and the time of arrival to obtain parametric equations of a circle, and the location of the source is the point of intersection. (b) The TDOA method uses the time difference between each pair of receivers to obtain parametric equations of a hyperbola, and the point of intersection of the hyperbolas is the location of the source. The uncertainty of a single measurement, shown by the thickness of a path, is governed by the cadence of the instrument. Spacecraft configuration determines how these areas overlap. If the overlapping is perpendicular, the area of uncertainty is minimised. If the overlapping is tangential, the area of uncertainty is maximised.}
   \label{Fig:TOAvsTDOA}
\end{figure}

\section{Methods}\label{sec:method}
The chosen method of localisation for this study is multilateration due to its simplicity.  Multilateration (also known as trilateration if only three receivers are used) uses timestamps from a number of different receivers to position the source of an emission using geometry. There are two main types of multilateration: Time-of-arrival (TOA) and time-difference-of-arrival, or TDOA, (see Fig.~\ref{Fig:TOAvsTDOA}). The main difference between these two methods lies in the time difference ($\Delta t$) required. The TOA method is the method of multilateration used by global positioning systems such as Galileo and GPS, and it uses a $\Delta t_{n0}$ between the emission time $t_0$ and the received time $t_n$, where $n= 1, 2, 3,...$, depending on the number of receivers. The TDOA method, on the other hand, does not require the time of emission $t_0$ but only the time received by the different receivers, and thus the $\Delta t_{ij}$ required is defined as $\Delta t_{ij} = t_i - t_j$, where $i$ and $j$ are subscripts for each of the receivers. This time difference ($\Delta t$) has implications in the geometrical analysis to detect the source of the emission where TOA is governed by the solution to the parametric equation of $n$ circles with radius
\begin{equation}
    d_n = c \ \Delta t = c \  (t_n - t_0) 
\hspace{1cm} \text{\textit{n = 1, 2, 3, ...}}
\end{equation}
and the equation of the circle
\begin{equation}
d_n^2 = (x_n - x )^2 + (y_n - y)^2 
\hspace{1cm} \text{\textit{n = 1, 2, 3, ...},}
\end{equation}
where $(x_n,y_n)$ corresponds to the coordinates of each of the spacecraft receivers. 
The TDOA method's geometrical approach is slightly different in that it requires solving the solution to the parametric equation of n number of hyperbolas as shown in the appendix of \cite{tri:badman2022tracking} and available at \cite{soft:samuel_badman_2023_10107890}. The disadvantage of this approach is that hyperbolas contain two branches, and therefore false positives may appear, but this can be mitigated by making further assumptions to choose one solution.
Due to the fact that TOA yields a single solution, as opposed to multiple ones; for its mathematical simplicity; and since it can be easily extended to utilise multiple spacecraft, we chose TOA for the purposes of this work.

\subsection{Bayesian multilateration}\label{sec:bayestri}
Bayesian statistics for multilateration is widely used in different fields, from communications where it is used to track Radio-frequency identification (RFID) nodes \citep{bay:zhou2009rfid,bay:sanpechuda2008review} to biology, where \cite{bay:reinwald2021seismic} used pressure waves and seismographs to track the movement of elephants in Africa. Here, we apply the same Bayesian concepts and techniques used by \cite{bay:reinwald2021seismic} and introduced by \cite{bay:speagle2019conceptual} to track solar radio bursts in the solar corona. 

The posterior of a Bayesian statistics positional problem is defined as the probability of finding the position of a source ($\vec{x}$) and the speed of signal propagation (v) given an observable $\Delta t$:
\begin{equation}
    P(\vec{x},v \ | \ \Delta t).
\end{equation}
According to Bayes' theorem, 
\begin{equation}\label{eq:bayes1}
    P(\vec{x},v \ | \ \Delta t) = P(\vec{x},v) \frac{P(\Delta t \ | \ \vec{x},v)}{ P(\Delta t)}.
\end{equation}
The conditional probability may be written in terms of the priors
\begin{equation}\label{eq:conditional}
    P(\vec{x} \ | \ v ) = \frac{P(\vec{x},v)}{P(v)}.
\end{equation}
Rearranging Eq.~\ref{eq:conditional} and substituting into Eq.~\ref{eq:bayes1} results in
\begin{equation}\label{eq:bayesianEQ}
    P(\vec{x},v \ | \ \Delta t) =  \frac{P(\Delta t \ | \ \vec{x},v) \  P(\vec{x} \ | \ v ) \ P(v)}{ P(\Delta t)},
\end{equation}
where each term is defined as: 
\begin{itemize}
    \item $P(\vec{x},v \ | \ \Delta t)$ is the posterior distribution or the probability of finding the source at position $\vec{x}$ with a speed of propagation $v$ given an observable $\Delta t$.
    \item $P(\Delta t \ | \ \vec{x},v)$ is the likelihood function or the probability of observing a $\Delta t$ given a position $\vec{x}$ and a speed of propagation $v$. This likelihood function is our physics model, and it is defined by the type of multilateration one wants to apply. In this case, TOA is governed by $\Delta t = d / v$, where  $d$ is the distance between the source and the receiver.
    \item $P(\vec{x} \ | \ v ) $ is the prior distribution of the source position given a speed of propagation $v$.
    \item $P( v ) $ is the prior distribution of the speed of propagation of the photons. The main assumption made by traditional multilateration methods such as TOA and TDOA is that the speed of propagation of the electromagnetic radiation is $c$, which is valid asymptotically far from the radio source but may break down close to the emission site where the emitted radiation is at a frequency similar to the local plasma frequency \citep[e.g.][]{Thejappa2010, Kontar2019}. The Bayesian solver does not make this assumption and allows for a distribution of the speed of propagation. The objective of this method is to accommodate the possibility that the ray from source to observer may have a longer than expected time of arrival due to scattering or refraction near the source but without needing to introduce detailed ray-tracing simulations \citep[e.g.][]{tri:musset2021simulations, tri:chen2023source}.  This can be modelled as a truncated normal distribution with a maximum of $c$ to restrict all solutions to physical solutions, or it can be modelled as a normal distribution with $c$ as the maximum, allowing for the Markov chain Monte Carlo (MCMC) solver (see Sect.~\ref{sec:bayesimplementation}) to test if the solution is physical or not. Convergence is expected at $v \approx c$ or $v<c$ for a physical solution. We note this implies that the path of the light ray is still along a straight line. 
    \item $P(\Delta t)$ is the evidence (i.e. observed) probability shown as a distribution of the time difference between the time of emission $t_0$ and the time of arrival $t_1$. The time of arrival is observed by the spacecraft, and the time of emission is fitted by the MCMC solver, which finds a time of emission that would be consistent for all spacecraft. 
\end{itemize}
\begin{figure*}
  \centering
  \includegraphics[width=0.6\textwidth]{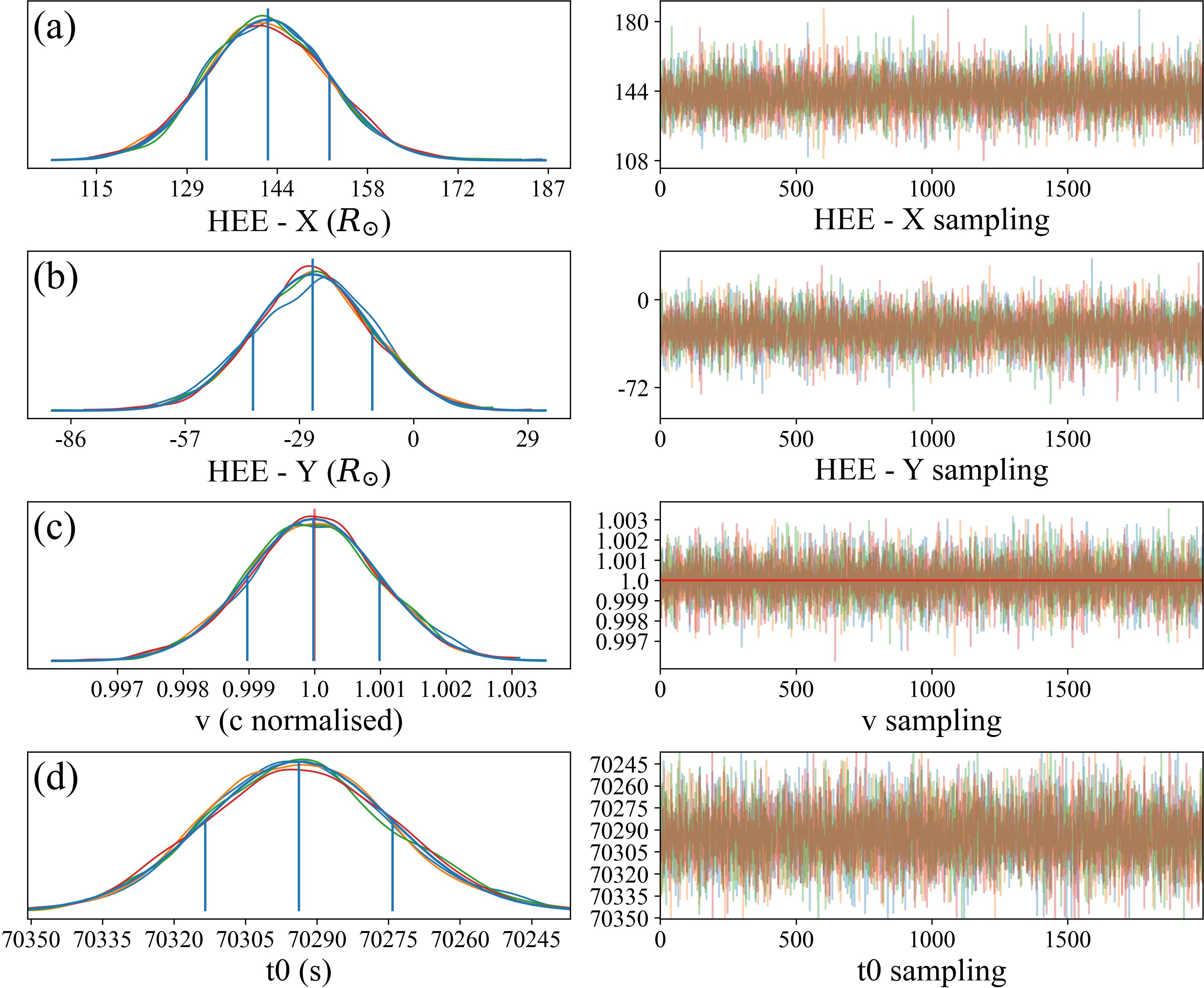}
  \caption{Example trace plots of the posterior distributions from BELLA at 0.154~MHz. Sampling chains are shown in the right column, while the corresponding distributions are shown in the left column. Vertical lines on the distributions show the peak and $1\sigma$ range of a combined distribution. Panels (a) and (b) are X and Y coordinate distributions, respectively, where the peaks are the coordinates of the source of the emission and $1\sigma$ shows the area of confidence. Panel (c) is the speed of propagation of the emission, v. The speed of propagation was allowed to be greater than c for the purposes of testing if the distribution converged at $v=c$ or $v<c$. A red line is displayed as a quick view of the $v=c$ limit. Panel (d) shows the time of emission t0 of the burst in seconds from midnight. The peak of the t0 distribution shows the most likely time of the emission, and $1\sigma$ shows the region of uncertainty for this time of emission.}
   \label{Fig:trace}
\end{figure*}
\begin{figure*}[!ptb]
    \centering
    \includegraphics[width=0.9\textwidth]{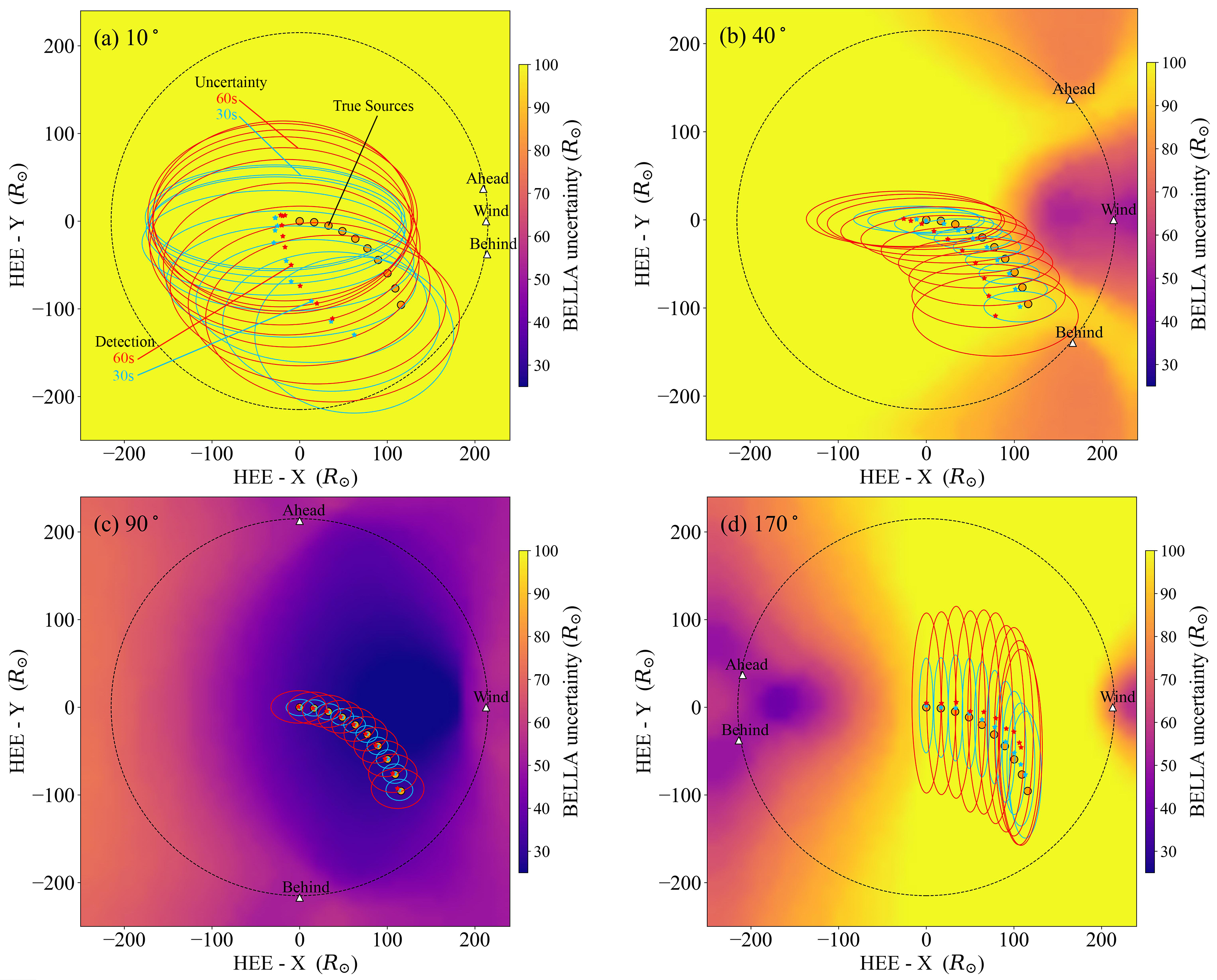}
    \caption{Simulated radio sources and their corresponding BELLA detection at (a) $\pm10^{\degr}$, (b) $\pm40^{\degr}$, (c) $\pm90^{\degr}$, and (d)$\pm170^{\degr}$ of Ahead-Behind separation with respect to Wind. Orange dots are the true location of the sources, and stars and ellipses are their corresponding detection and uncertainties, respectively. At each spacecraft configuration, simulations were performed twice with different cadences: 30 s (blue) and 60 s (red). The background maps were generated at a 60-second cadence and show a general overview of the multilateration uncertainty. Uncertainty of the measurements is dominated by the cadence of the instruments and the location of the spacecraft. The best performance of the multilateration technique occurs when the separation angles between all spacecraft are maximised and widely distributed.}
    \label{fig:Simulation_frames}
\end{figure*}
\begin{figure}
    \centering
    \includegraphics[width=8cm]{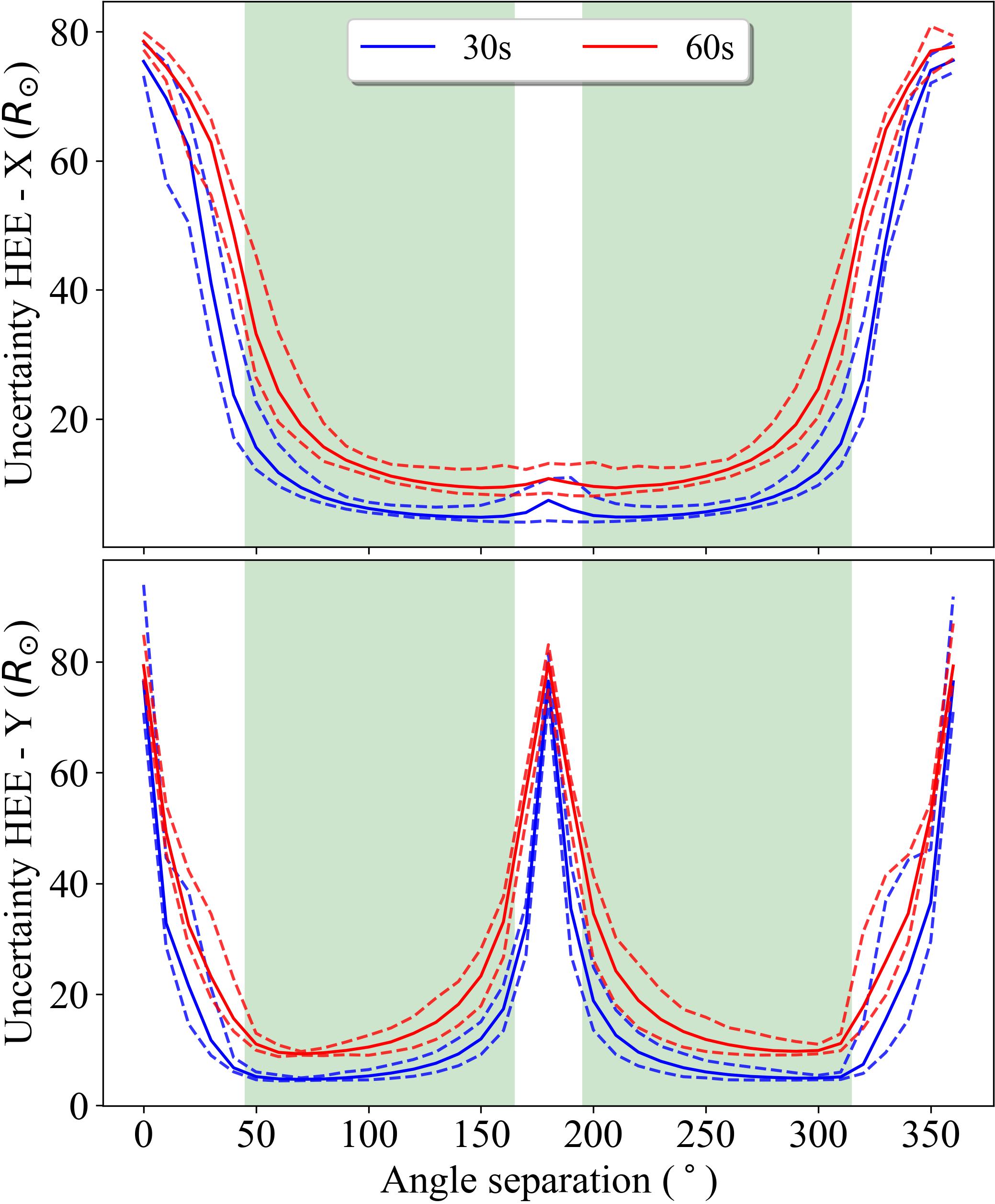}
    \caption{Simulation showing the expected uncertainties when performing multilateration on a STEREO and Wind system. The solid line represents the average $1\sigma$ uncertainty of the sources in the simulated burst, and the dashed line represents the maximum and minimum $1\sigma$ uncertainties of the positions of the simulated bursts. The angle of separation of the Ahead and Behind spacecraft of the STEREO system with respect to Wind shows a strong contribution on the uncertainty of the localisation method when two or more spacecraft are close to each other. This simulation was performed assuming different cadences, which also had an effect on the performance of the multilateration, but it was not as dominant as the angle of separation. These simulations suggest that BELLA will have a minimal contribution to the uncertainty of the results if the angle of separation is in the range of $\pm [45\degr,165\degr]$. This range is defined as the `sweet spot' region and is highlighted in green. See Fig.~\ref{fig:Simulation_frames} for frames of this simulation. }
    \label{fig:SimulationSummary}
\end{figure}

\subsection{Implementation of Bayesian multilateration}\label{sec:bayesimplementation}
The Bayesian multilateration (see Sect.~\ref{sec:bayestri}) was performed with the PyMC \citep{other:salvatier2016probabilistic,other:pymc_v5.6.1} package in Python. The PyMC package allows for Bayesian inference based on probabilistic models and creates an environment where prior distributions can be defined intuitively within a model container that performs Bayesian statistics automatically.  
BELLA makes use of the normal distribution function embedded in PyMC as its primary source of prior distributions. The probability density function (PDF) of a normal distribution is
\begin{equation}\label{eq:normalPDF}
    f(x \ | \ \mu, \sigma) = \sqrt{\frac{1}{2\pi\sigma^2}}\exp{\bigg(-\frac{1}{2\sigma^2}(x-\mu)^2\bigg)},
\end{equation}
where $\sigma$ is the standard deviation and $\mu$ is the mean.

In order to perform Bayesian inference, a PyMC model context manager was created. The priors were then defined as follows:
\begin{itemize}
    \item The prior for the coordinate space of the source positions $P(\vec{x} \ | \ v ) $ was implemented as a stochastic random variable sampled from a normal distribution with $\mu=0$ and $\sigma=h R_{\odot}$, where $h$ is an educated guess of the radial distance of the emission to the Sun in the range [-215, 215] $ R_{\odot}$. This guess does not affect the posterior distribution, but it does aid in achieving convergence. The term $h$ was chosen to be one-fourth of the space domain in one dimension, which is $\sim$ 80~$ R_{\odot}$. 
    \item $P( v )$ is the speed of propagation prior defined as a stochastic random variable sampled from a normal distribution with $\mu=c$ and $\sigma=0.1c$.
    \item $P(\Delta t)$ is the probabilistic evidence implemented as a deterministic variable obtained by sampling from a $t_0$ uniform distribution with limits much larger than the light travel time corresponding to the coordinate space limits (for example, $t_{lims}=\pm$24~x~60~x~60~s) and subtracting $t_0$ from the expected time of arrival $\Delta t = d / v$, where $d$ is the distance between the location of the spacecraft $\vec{x_{sc}}$ and the prior for source positions $\vec{x}$ as $d = \lvert  \vec{x_{sc}} - \vec{x}\rvert $,  and $v$ is sampled from the speed of propagation prior. 
\end{itemize}
We note that the speed of propagation in the model is not $c$ but is determined by the distribution of propagation speeds $P( v )$. This is one of the advantages of this method, as it no longer assumes that the speed of propagation is $v=c$. We also emphasise that the prior $P( v )$ was allowed to offer sampling where $v>c$. This was done for the purpose of testing if the Bayesian solver converges to a propagation speed that is physical (i.e. $v\leq c$ ). Thus, if the maximum of the posterior distribution shows that $v> c$, the results are no longer considered physical, and further investigation is necessary. Another benefit of this method is that it is fully compatible with any number of spacecraft as long as $n\geq 3$ (and without the need to alter the code).

Once the priors are defined in the PyMC model context manager, the likelihood function $P(\Delta t \ | \ \vec{x},v)$ is defined as an observed stochastic variable that samples from a normal distribution with $\mu=\Delta t$ (calculated earlier) and $\sigma=t_{cadence}$ (instrument cadence). Then, an observed parameter $t_{toa}$ (the time of arrival as measured by each spacecraft at a given frequency) is used by the PyMC environment for validation. 

With the parameter space set up, the PyMC environment was then ready for sampling. The PyMC package uses by default the No-U-Turn Sampler \citep[NUTS;][]{other:hoffman2014no}, which is a self-tuning sampling method that prevents the sampling from retracing its own steps and thus helps efficiently explore the target distribution. 
The sampler was set up in the PyMC context manager to perform four different sequences of samples, also known as chains. Each chain was initialised using 2000 tuning samples, and then an additional 2000 samples were drawn to achieve a singular stable solution, also known as convergence. When convergence was achieved, the posterior distributions ($P(\vec{x},v \ | \ \Delta t)$) were found to be approximate and discrete normal distributions. The mean of the distributions could then be defined as $\mu=[x_x, x_y]$. This is the most probable source location with $1\sigma$ uncertainty for each coordinate. The area of confidence was obtained by generating an ellipse with the axes $\sigma=[\Delta x_x,\Delta x_y]$ as read off the resulting distributions. 

Figure~\ref{Fig:trace} shows an example of the trace plots generated after one multilateration. Each of the colours in the figure represents a different chain. The columns in the figure represent the exact same data: the left column is the distribution of the sampled values, and the right column is the value of the sample values as a function of index. From top to bottom, we found the posterior distribution of the x coordinate, the posterior distribution of the y coordinate, the distribution of the speed of propagation, and the distribution of the time of emission with respect to the observed time. Throughout this work, we use the peak of these distributions as the most probable location of the source and 1$\sigma$ as the uncertainty of the distribution. The peak and $\sigma$ values are highlighted with vertical lines, as seen in Fig.~\ref{Fig:trace} (left). The distribution of the speed of propagation also displays a red line corresponding to the $v=c$ limit. This was used as a quick view warning to test the $v\leq c$ criterion. 

\section{Simulations and method validation}\label{sec:bayessimulation}
In order to first validate the method and its capabilities, a simulation was set up under the controlled conditions of a known set of source positions and a known propagation speed (i.e. no refractive or scattering effects). In these simulations, we recovered the ground truth signal to a better or worse extent depending on the spacecraft configuration. This allowed us to test for the contribution of the spacecraft configuration to the uncertainty of the localisation results. Figure~\ref{fig:Simulation_frames} and Fig.~\ref{fig:SimulationSummary} show the results of these simulations, with the first figure showing four different instances of the simulation and the latter providing a summary. In order to see the effects of instrumental cadence on the performance of the simulations, all cases were simulated twice: once at a cadence of 30\,s (blue) and once at a cadence of 60\,s (red). The background maps were all simulated at 60-s cadence. In addition to validating the methodology, these simulations allowed for the selection of a suitable Type III SRB candidate by discarding spacecraft configurations that would guarantee uncertain results. The simulation was composed of the following: A three-spacecraft system analogous to the real STEREO A/B and Wind configuration was established. For simplicity, the Ahead and Behind spacecraft were kept at a constant distance of $\pm1\%$ of 1~au, respectively. The Wind spacecraft was kept stationary at L1, while the Ahead and Behind spacecraft were synchronously separated from Wind in $\pm10\degr$ increments, respectively. 
In order to investigate the effect of spacecraft geometry and instrument time resolution on the inherent uncertainty of the method, a mesh-grid of $500$~x~$500$ nodes was established, and each node was treated as an emission source. By assuming free streaming and a propagation speed of $v=c$, synthetic arrival times were calculated as $t = d/c$ for each source-spacecraft pair. A small amount of Gaussian noise (on the order of <1\%) was applied to the data. If there were regions where the uncertainty was above a user-specified threshold despite applying the smallest amount of noise, then the configuration was rejected. The user-specified threshold was 25~$R_\odot$, as this is the light travel distance for an instrument with a 60-s cadence (typical instrument time resolution). The noisy data was used as an input for the Bayesian procedure described in Sect.~\ref{sec:bayesimplementation}. Posterior distributions for the detected position at each node gave corresponding $1\sigma$ uncertainties in the $x$ and $y$ directions. The maximum value of the $x$ and $y$ uncertainties was taken as the value for the uncertainty map colour. As shown in Fig.~\ref{fig:Simulation_frames}, it is clear that when two or more spacecraft are close together, a source anywhere in the ecliptic plane has a large uncertainty (yellow regions), but when they are equidistantly spaced, the uncertainty is minimised (blue regions).
A test burst was also simulated by selecting a number of simulated sources along a notional Parker spiral and performing the same procedure at two different cadences (60~s and 30~s). By recovering this known ground truth, we validated the method, and the validation allowed for testing of the effects of the instrument cadence on the performance of the multilateration. The test burst also allowed for a visual depiction of the discrepancy between the multilaterated source (shown as red and blue stars) and its true counterpart (orange circles) as well as showing the difference in the $x$ and $y$ uncertainties.
Figure~\ref{fig:SimulationSummary} shows the average uncertainties for the full test burst in solid lines, and the maximum and minimum uncertainties of the nodes are shown as dashed lines. The `sweet spot' regions are highlighted in green and are the regions where both the $x$ and $y$ uncertainties are under a user-defined threshold. Moreover, these regions served as a starting point for finding suitable candidates. The threshold in Fig.~\ref{fig:SimulationSummary} was defined as the largest of the midpoints of the decay and rise in uncertainty regions of the 60-s cadence simulation. This corresponds to the $\pm[45\degr,165\degr]$ bands. Figure~\ref{fig:SimulationSummary} also shows that for the ideal case of nearly equidistant spacecraft, the ground truth is retrieved accurately, meaning at the expected uncertainty due to the light travel time corresponding to the cadence. However, as $n$ number of spacecraft move closer together, they start to behave as a unique receiver. This is particularly evident for the $180\degr$ case, where the Ahead and Behind spacecraft are aligned with Wind. In this particular case, the system becomes one dimensional, and the $X$ coordinate uncertainty remains stable as a minimum, while the  $Y$ coordinate becomes unstable, and the uncertainty becomes a maximum. The reason for this is that in a one-dimensional system, only two spacecraft are needed to multilaterate a source, and therefore the $X$ coordinate still meets the minimum requirements for multilateration. 

\section{Observations}\label{sec:observation} 

\begin{figure*}
\centering
   \includegraphics[width=\textwidth]{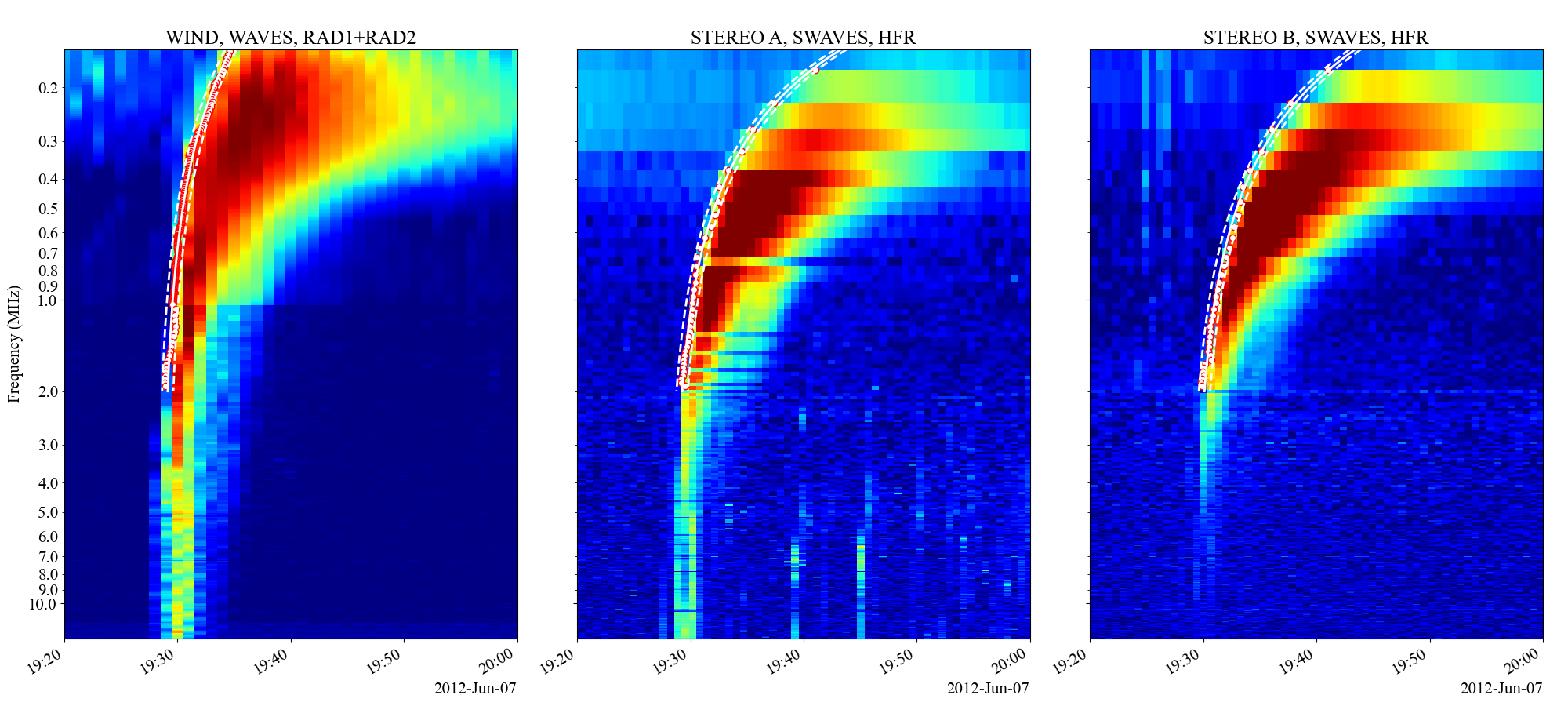}
     \caption{Type III radio burst observed by Wind (left), STEREO A (middle), and STEREO B (right) on 7 June 2012. The event was chosen due to its relative brightness with respect to the background. The burst was observed to be an isolated event and lasted approximately 20 mins. The frequency range of the burst located by each of the spacecraft was from 13 MHz to 0.1 MHz; however, STEREO B showed a poor S/N at the higher frequency range, presumably due to solar disk occultation. The white line is the fit of the Type IIIs used for the multilateration in the range of 2 MHz to 0.15MHz.}
     \label{fig:DynSpecHoriz}
\end{figure*}
\begin{figure}
\centering
   \includegraphics[width=0.4\textwidth]{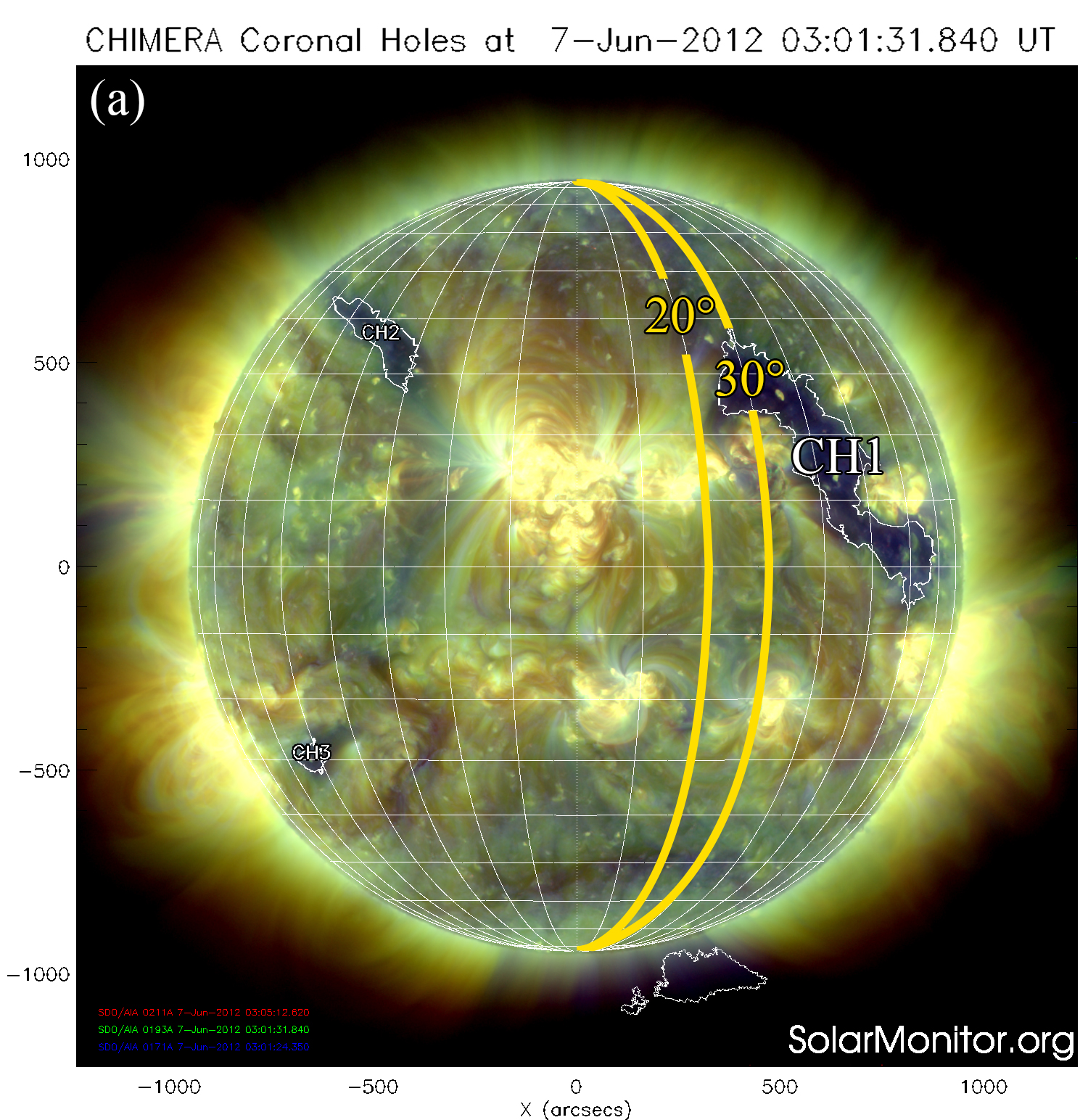}   \includegraphics[width=0.4\textwidth]{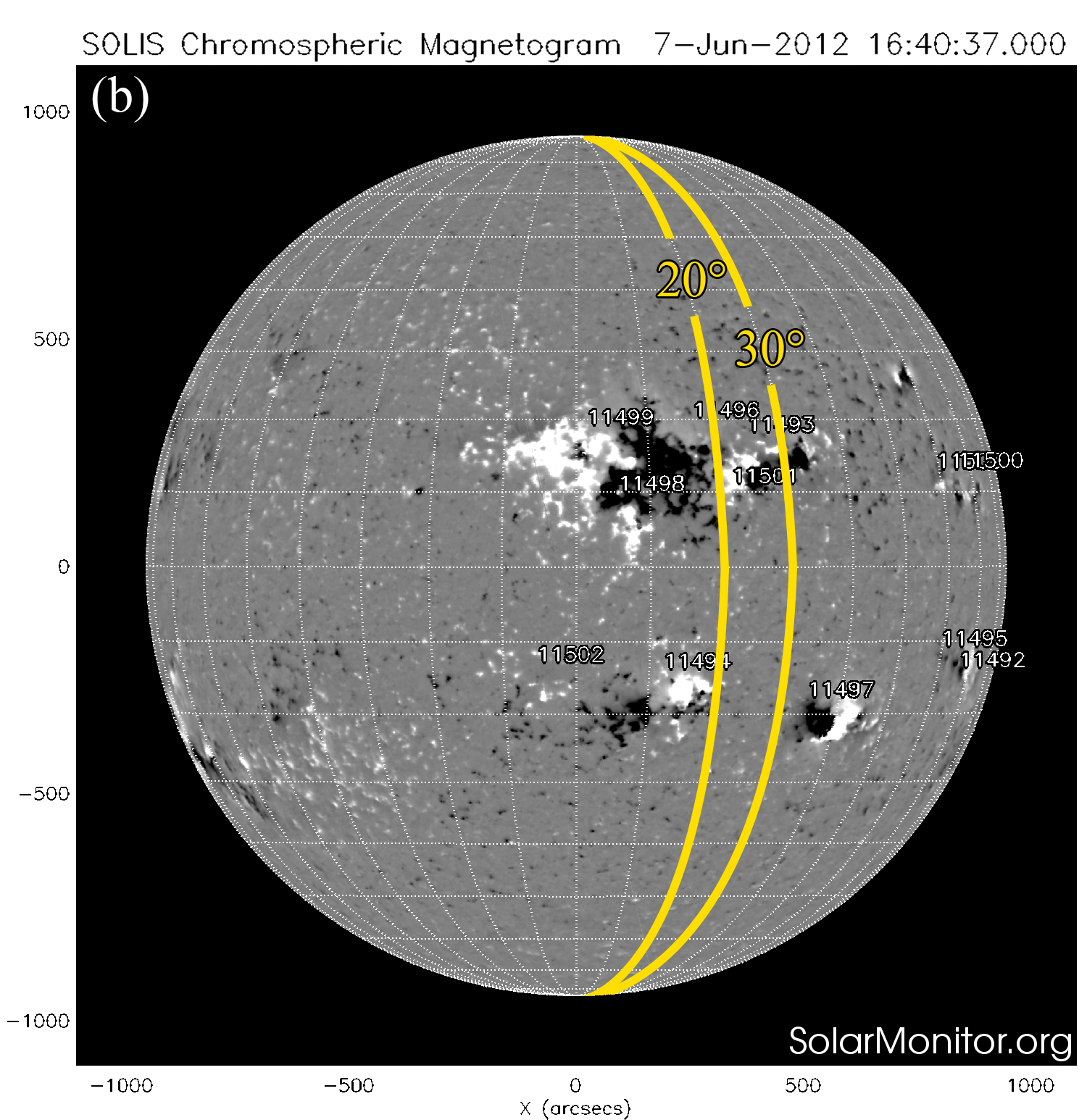}
     \caption{Coronal holes on 7 June 2012 obtained by CHIMERA. Fast solar wind regions observed in Fig.~\ref{fig:resultsHUXT} are consistent with the location of CH1. Magnetogram data from SOLIS show 10+ as the number of NOAA active regions, four of which are adjacent to the East of CH1. This is later shown to be consistent with the location of the triangulated Type III beam. 
     }
     \label{fig:solaractivity}
\end{figure}

To validate our method and test it on a real event, we examined in detail an individual Type III burst that was clearly observed by multiple well-separated spacecraft (see Fig.~\ref{fig:DynSpecHoriz}). The event chosen for this study occurred on 7 June 2012 from 19:30 UT and was observed by Wind  \citep[Wind/Waves,][]{inst:Bougeret1995waves}, STEREO A, and STEREO B \citep[STEREO/Waves,][]{inst:Bougeret2008swaves}). The datasets were obtained from \cite{data:WindL2} and \cite{data:L3STEREOGP}. During the date of the event, the coronal hole identification via a multi-thermal emission recognition algorithm \citep[CHIMERA,][]{other:garton2018automated} reported a large coronal hole in the north-western quadrant of the Sun spanning $\sim$~$30\degr$, labelled as CH1 in  Fig.~\ref{fig:solaractivity}~(a). Figure~\ref{fig:solaractivity}~(b) shows over 10 active regions as observed by the SOLIS Vector Spectro Magnetograph\cite[SOLIS VSM;][]{inst:henney2006solis}) and labelled by the National Oceanic and Atmospheric Administration (NOAA). This makes this region a highly plausible Type III source region given the proximity of open field lines to active region loops. As we show, BELLA allows for a strong inference to back up this hypothesis for our studied burst. 

The Type III radio burst was observed by Wind from 0.06 to 13.825 MHz, by STEREO A from 0.15 to 16 MHz, and by STEREO B at a range similar to that of STEREO A but with signs of solar disk occultation, as the higher frequencies of the burst show a decay in brightness. This particular event was chosen for a number of reasons. The location of the spacecraft is optimal for multilateration techniques. The STEREO spacecraft are at $\pm 116\degr$ from the Wind spacecraft, which resulted in near-equidistant separation between all three spacecraft. At this separation, Sect.~\ref{sec:bayessimulation} showed that the intrinsic uncertainty from BELLA is minimal and expected to be dominated by the cadence of the instrument and physical errors.
The Type III radio burst is isolated. There is no other bursts or activity that could potentially cause confusion or uncertainty to the edge of the burst. 
The burst is bright, and the edges are well defined. The low cadence of the instruments was a limiting factor, and therefore, the burst picked for the validation of BELLA was chosen upon visual inspection to be as clean as possible and with as high S/N as it was possible. 
All spacecraft are at $\sim$1~au from the Sun, simplifying the system analysed. It is yet unknown whether the effects of spacecraft distances from the Sun has an effect on the performance of the algorithm. As it is beyond the scope of this initial test of BELLA, we will explore how spacecraft heights, and other properties, can affect the performance of the algorithm in future studies.

\begin{figure}
\centering
    \includegraphics[trim={0 3cm 0 0},clip,width=10cm]{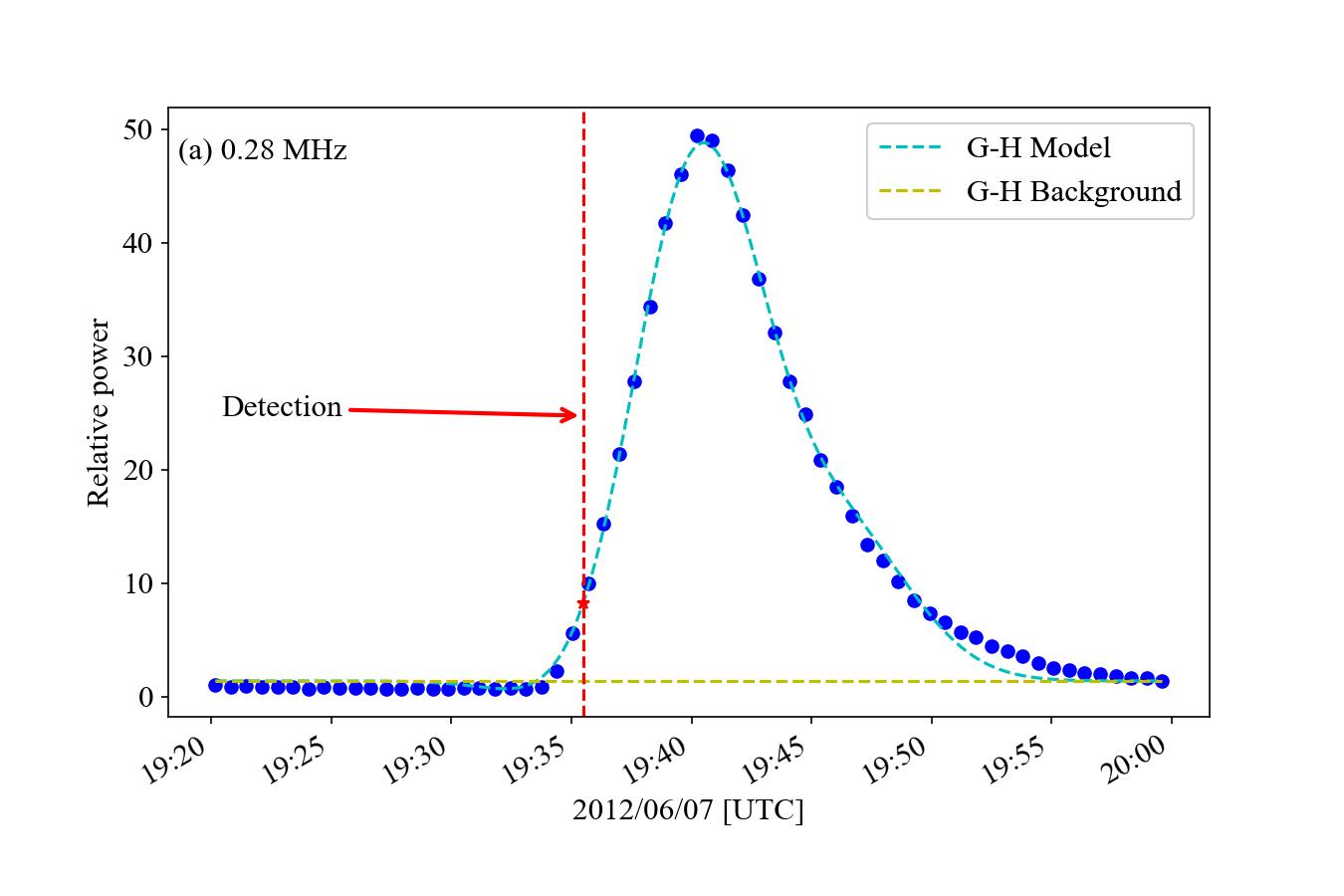}
    \includegraphics[trim={0 3cm 0 1.6cm},clip,width=10cm]{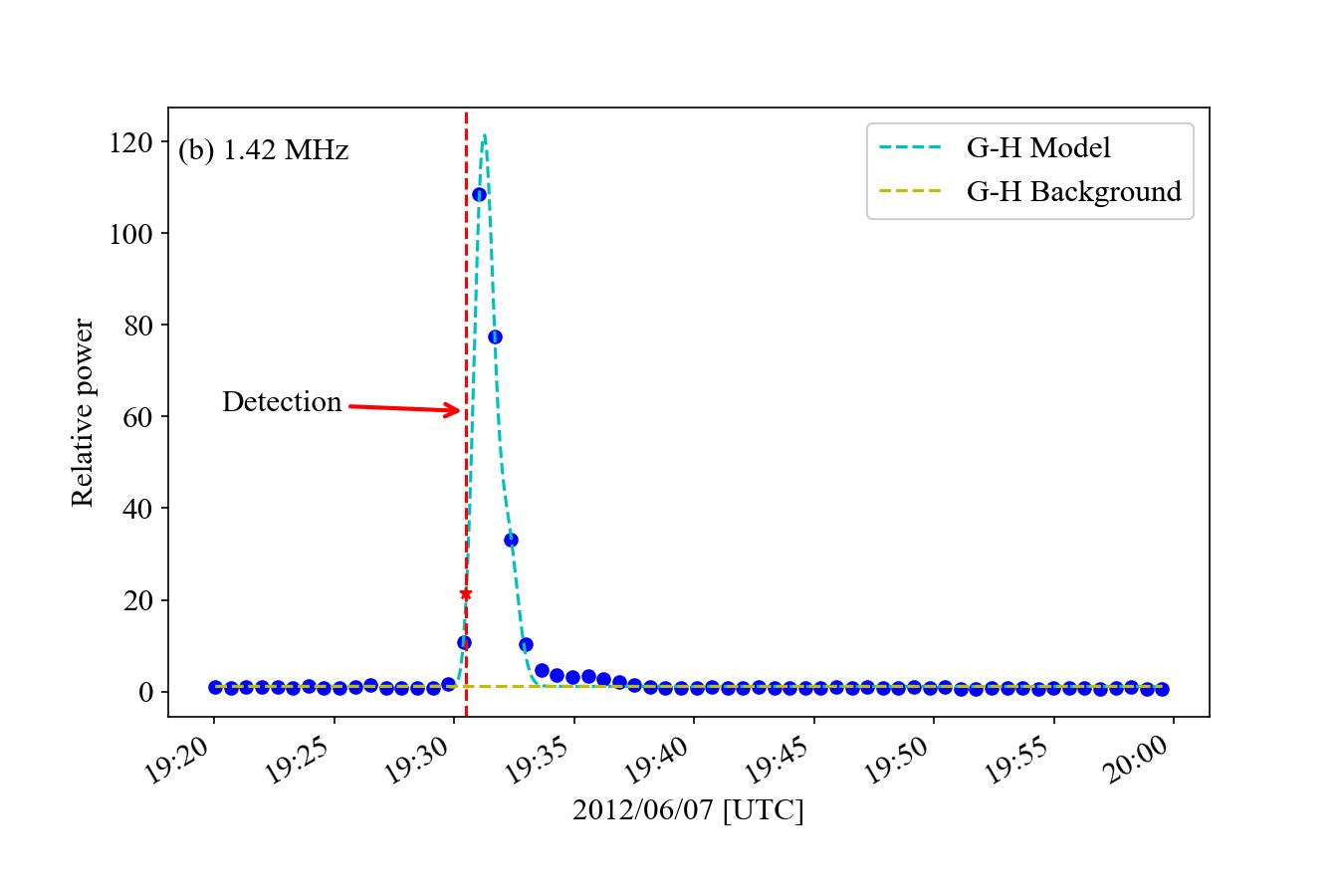}
    \includegraphics[trim={0 0 0 1.6cm},clip,width=10cm]{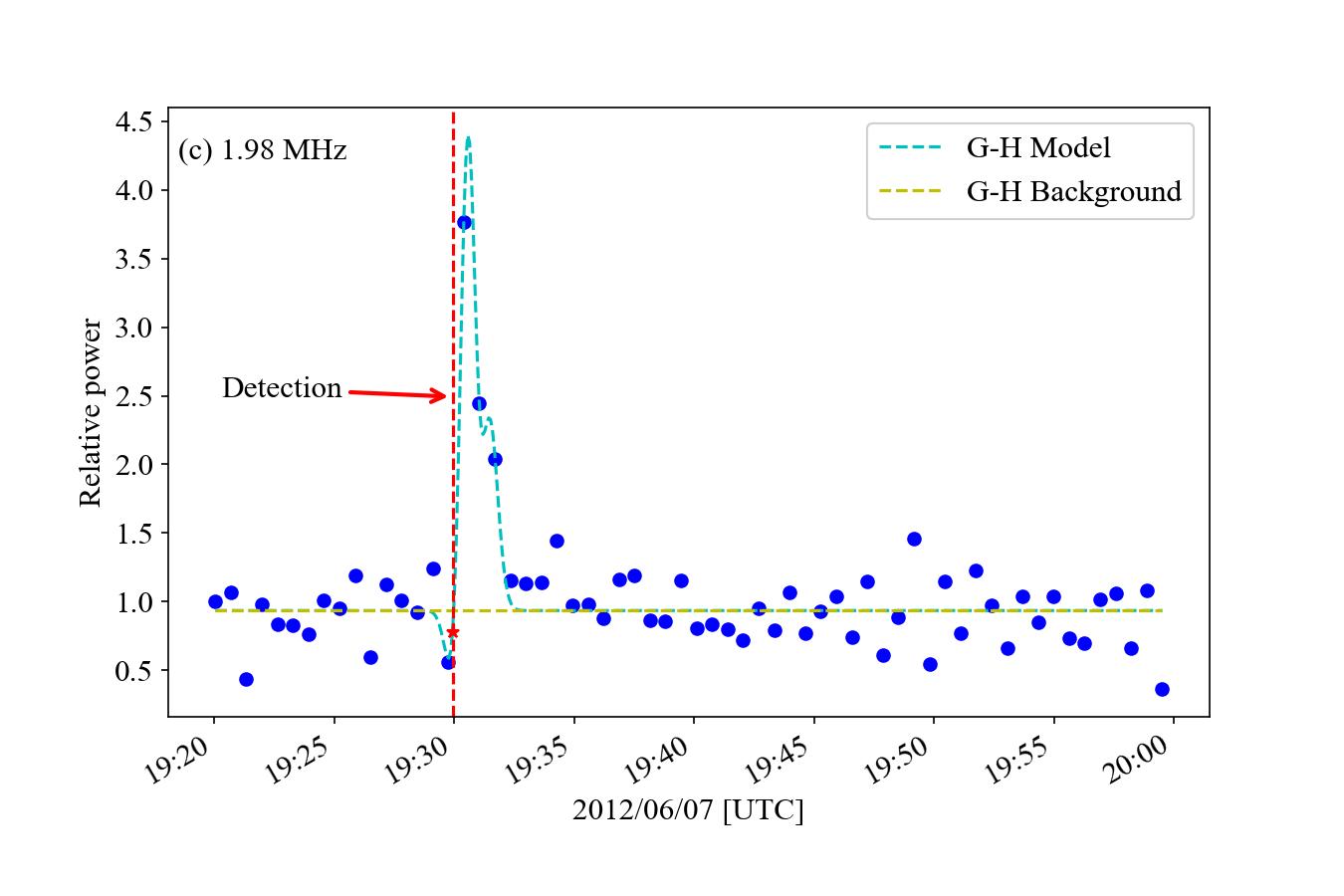}
    \caption{Edge detection from the GH Type III detection algorithm. Blue dots show the light curve data, the yellow dashed line shows the background level derived from the GH Type III fitter, the cyan dashed line shows the GH fitting, and the red star and dashed line show the detection point used to obtain the rise time of the burst. a) Example of a high S/N light curve fit using GH polynomials. b) Example of a fit with a low signal data count. The low data count had no noticeable effect on the performance of the algorithm to obtain the rise time of the light curve. c) Example of a light curve with a poor S/N. The GH algorithm was capable of fitting a Type III light curve profile despite the poor S/N. The detection was observed to be on the order of one cadence from the rise time location based on visual confirmation of the data points in the dynamic spectra plot in Fig.~\ref{fig:DynSpecHoriz}.}
     \label{fig:GHfitting}
\end{figure}

The leading edge of the Type IIIs was chosen as the determining feature to obtain time measurements for the multilateration, as it corresponds to the shortest possible path of the radiation from the source to the receiver. Determining a consistent point for the rising time of a Type III light curve for a particular frequency poses a number of challenges when the time resolution of an instrument is as low as 60~s, which is the case of the Wind/Waves instrument, and 38.05~s, as in the case of STEREO/Waves. In order to obtain a consistent rise time for the light curve of a Type III, we fit a Gaussian-Hermite (GH) model \citep{fit:van1993new} from the Python package Kapteyn \citep{fit:KapteynPackage}. The GH model is an asymmetric Gaussian-like function that derives values for the skewness and kurtosis of the PDF. Figure~\ref{fig:GHfitting} shows different examples of the result of fitting a GH model using the {\it kmpfit} module of the package for a number of different light curves at varying frequencies. Figure~\ref{fig:GHfitting}~(a) is an example of a low-frequency light curve, and Fig.~\ref{fig:GHfitting}~(b) is an example of a high-frequency light curve. Figure~\ref{fig:GHfitting}~(c) is an example of a light curve with a poor S/N. The criteria for the rise time of the light curve was defined as $1\sigma$ to the left of the skewed distribution obtained from the GH fitting parameters.

Time stamps for the multilateration were extracted from the exact same frequencies. In order to account for the difference in frequency channels between STEREO/Waves and Wind/Waves as well as the low temporal resolution of these instruments, we fit a time evolution function to the rise times detected using the GH method. This time evolution function was parameterised using a polynomial of the form
\begin{equation}\label{eq:timeevo}
    t(f) = a_2\frac{1}{f^2} + a_1\frac{1}{f} + a_0,
\end{equation}
where $t$ is the rise time; $f$ is the frequency; and $a_2$, $a_1$, and $a_0$ are coefficients that represent the curvature, the drift rate, and initial position of the burst, respectively. These fits can be observed in Fig.~\ref{fig:DynSpecHoriz} as white lines and show the data extracted for the multilateration. The white dashed lines represent the edges from where the MCMC sampled for the time prior of the Bayesian calculation (see Sect.~\ref{sec:bayesimplementation}). In addition, we directly interpolated individual rise times in order to assess the impact of the time evolution fitting (see Fig.~\ref{fig:DynSpecHoriz}).


\section{Results and discussion}\label{sec:resultsanddiscussion}
\begin{figure*}
\centering
   \includegraphics[trim={0 0cm 0 0},clip,width=\textwidth]{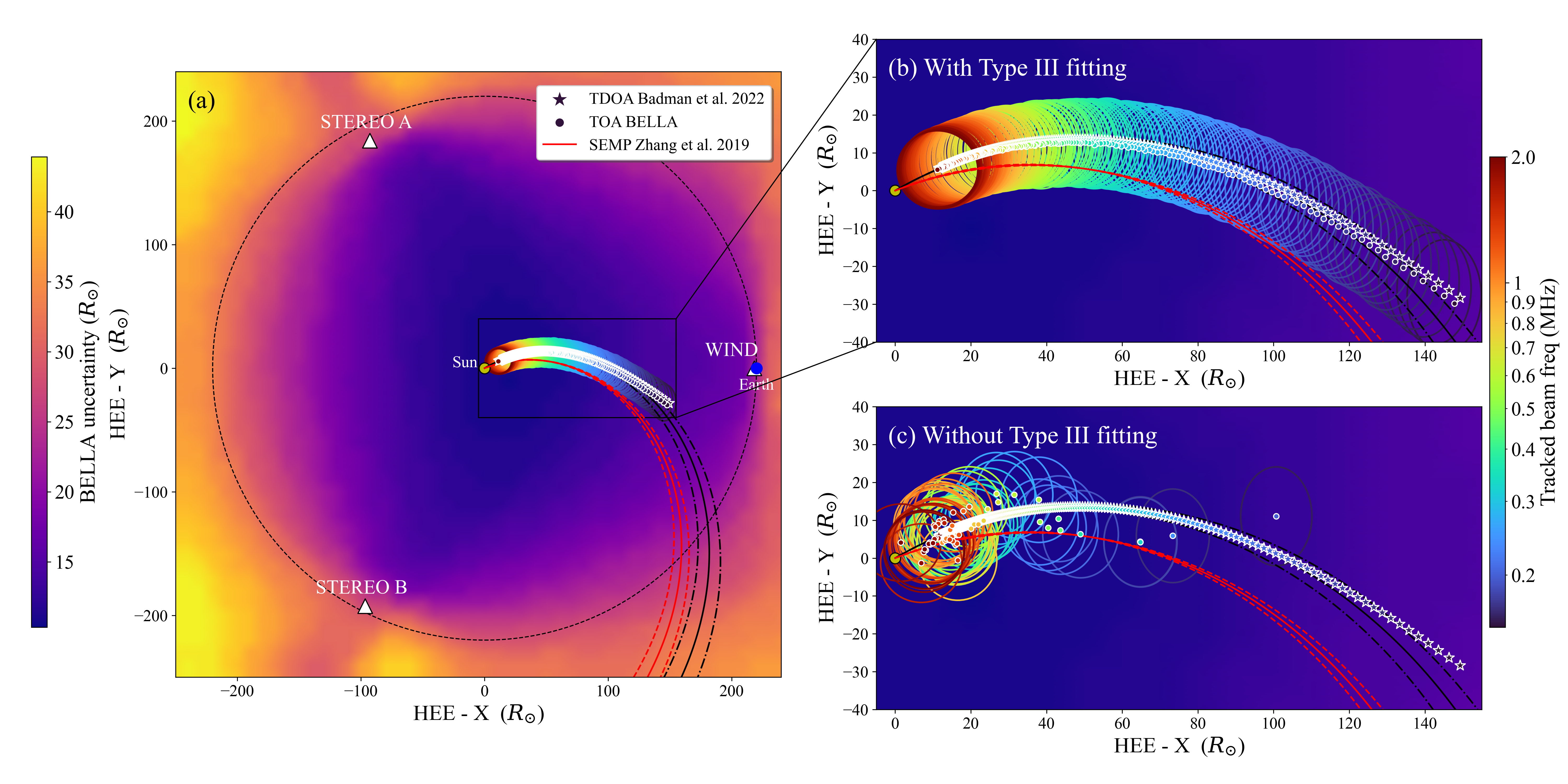}
     \caption{BELLA multilateration of the Type III SRB. (a) Top view of the ecliptic plane showing the location of the spacecraft and the multilaterated burst. A background map generated by simulated data tests the performance of BELLA for this particular spacecraft configuration at the location of the sources; (b) zoom-in of the burst; (c) same zoom-in, but the multilateration was performed on the output of the GH polynomial without the final fit of the Type III. We found that BELLA showed results with a confidence area of $\sim$~$15 - 20~R_\odot$ in diameter. The TDOA results from the method in \cite{tri:badman2022tracking} and SEMP have also been included for the purpose of validating the results obtained by BELLA. Parker spirals show the results from fitted BELLA points (black) and SEMP (red) at 400~km~s$^{-1}$, with dashed lines indicating the uncertainty at $\pm20$~km~s$^{-1}$. We found that there are some systematic differences between TDOA/BELLA ($\phi_0 \sim 30\degr$) and SEMP ($\phi_0 \sim 20\degr$), but they are consistent within the error bars that BELLA allows to be quantified.}

    \label{fig:resultsBELLA}
\end{figure*}

\begin{figure*}[!]
    \centering
    \includegraphics[trim={0 0cm 0 3cm}, clip, width=0.65\textwidth]{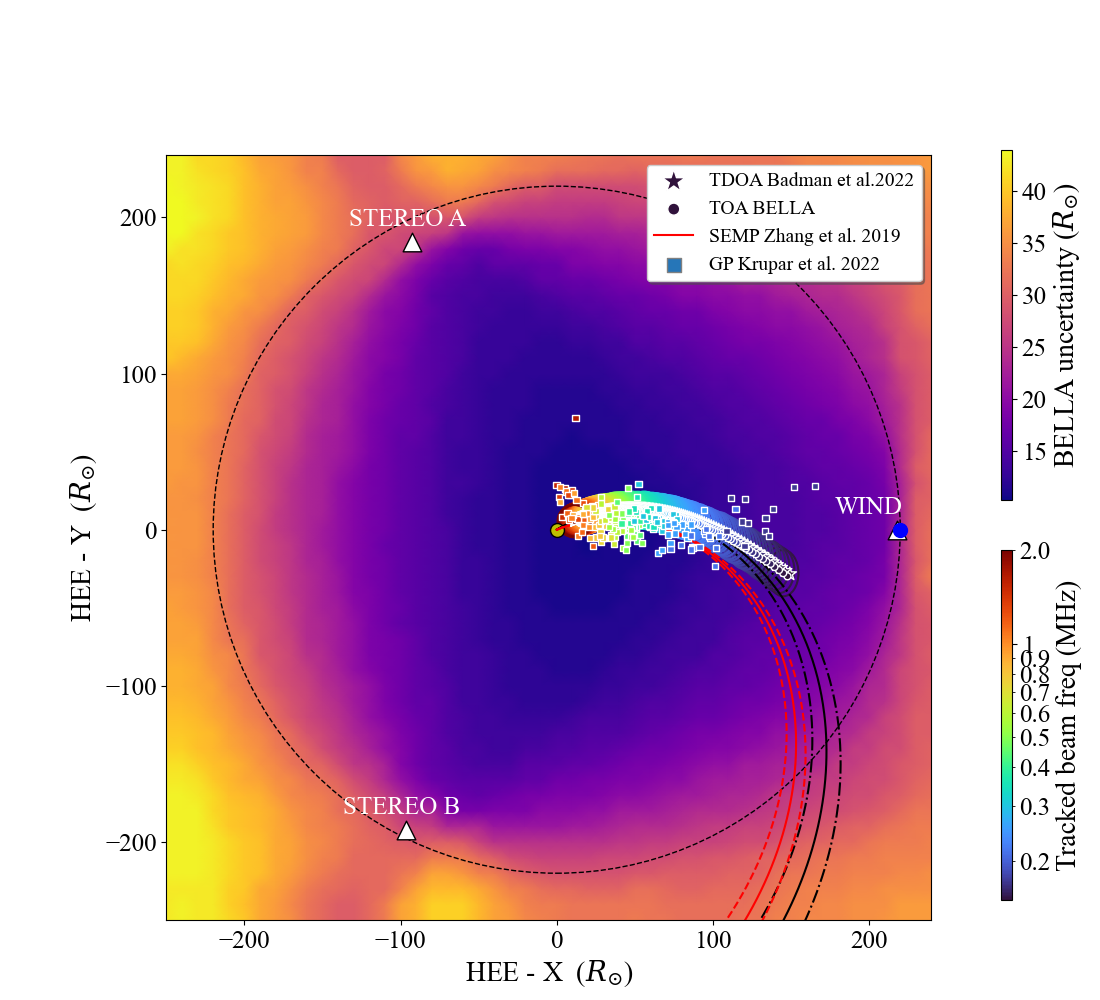}
    \caption{STEREO A/B L3 GP inversions \citep{data:L3STEREOGP} overlaid on Fig.~\ref{fig:resultsBELLA}~(a). The GP points show a clear trend in agreement with BELLA in both orientation and frequency distribution (as indicated by the colour scales). The GP results have a larger spread and make interpretation of an underlying spiral more difficult. All localisation methods are shown to be in agreement, illustrating that complementary but independent methods give approximately the same answer, which further validates the BELLA methodology.}
    \label{fig:GPvsBELLA}
\end{figure*}

The results of BELLA are shown in Fig.~\ref{fig:resultsBELLA}. The background map was obtained by simulating emission from every pixel in the map, and it displays the contribution of BELLA to the uncertainty of the results. The blue regions of the background map are areas where BELLA has a minimal effect in the uncertainty of the output positions and the yellow regions are regions where BELLA cannot accurately position a source, giving results similar to Fig.~\ref{fig:Simulation_frames}~(d). The large trianguloid shape of the blue region is a direct consequence of the optimal position of the spacecraft for that particular date. The spacecraft were equally spaced, and therefore a symmetrical output was expected. Simulations of a similar burst with varying spacecraft positions shown in Figs.~\ref{fig:Simulation_frames} and \ref{fig:SimulationSummary} show that the contribution of BELLA to the uncertainty of the measurements due to the position of the spacecraft is negligible for the position of the Ahead and Behind spacecraft during this event.\\
The map was generated using simulated data with a cadence of 38~s, which corresponds to a light travel time uncertainty of $\sim$~$15\,R_\odot$. This simulation differs from that in Fig.~\ref{fig:Simulation_frames}, as the spacecraft were chosen directly from their heliographic locations on the date and time of the burst. From the map, we observed that the whole burst occurred in a region of $<$~$15\,R_\odot$ in uncertainty. This means that BELLA has a negligible contribution to the uncertainty of the results, and any source of uncertainty is dominated by the cadence of the instruments and other (mostly physical) sources of uncertainty. This is consistent with the dynamic spacecraft simulation shown in Fig.~\ref{fig:Simulation_frames}.\\ 
Overlaid on this map is the BELLA output from the TOA multilateration performed on the Type III radio burst characterisation from Fig.~\ref{fig:DynSpecHoriz}. The burst is colour-coded by the frequency of emission (right-hand colour bar). The ellipses around each of the source points are the 1$\sigma$ uncertainties obtained from the BELLA posterior distributions. Also overlaid on this map is the output from the TDOA multilateration using the method from \cite{tri:badman2022tracking}, and shown in red is the output from the Solar radio burst Electron Motion Tracker \citep[SEMP,][]{tri:zhang2019forward} at 400~km~s$^{-1}$ $\pm \,20$~km~s$^{-1}$  solar wind speed. The three methods of positioning were found to be in agreement with each other within 1$\sigma$ uncertainty. We observed that the TDOA method yielded results nearly identical to the TOA method used by BELLA. This is in agreement with \cite{tri:kaune2012accuracy}, which showed that Monte Carlo simulations of TOA and TDOA have no discernible results in accuracy. The results indicate that the TOA solution converged to the same solution that TDOA produces geometrically and that the posteriori signal propagation distribution is close to $v=c$. We note that the line of sight of the radio waves propagation is for the most part unobstructed by the lower layers of the solar corona, which makes this burst ideal for the assumption of $v=c$. The purpose of this study is to validate BELLA as a method for positioning the source of observed radio bursts. This convergence yields two possibilities: a) the effects of refraction and scattering can be neglected for this particular burst or b), in this case, BELLA has converged on an `apparent' position of the burst shifted by propagation effects \citep{tri:chen2023source, scat:kontar2023anisotropic}. The effects of refraction and scattering will be the subject of further investigation, but as discussed later in this section, we find evidence that (b) is more likely for this burst. \\
Additionally, Level 3 (L3) GP inversions from STEREO data available at \cite{data:L3STEREOGP} are shown in Fig.\,\ref{fig:GPvsBELLA}. The GP triangulation was performed using the time of peak flux of the Type III as opposed to the leading edge because the GP inversions at the leading edge are not stable. BELLA, on the other hand, is designed to utilise the leading edge of the Type III, making the two independent methods complementary to each other. A clear trend was observed between the results obtained by the triangulation and by the multilateration, with only a small number of points lying outside of the $1\sigma$ area of uncertainty. The agreement in orientation and frequency distribution along the spatial domain serves to validate BELLA as a suitable method of localisation. \\ 
Figure~\ref{fig:resultsBELLA} also shows the results from SEMP, which assumed a pre-specified constant solar wind speed along the Parker spiral. The SEMP longitude $\phi_0$ parameter was found to be $\sim$~$\phi_0=20\degr$. A Parker spiral, shown as a black dotted line in the figure, was also overlaid on the map. The parameters for this spiral were obtained by assuming a Parker spiral to be an Archimedean spiral:
\begin{equation}\label{eqn:spiral}
    r(\phi) = \frac{v_{sw}}{\Omega_\odot}(\phi - \phi_0) + r_0,
\end{equation}
where $r$ is the distance from Sun, $r_0$ is the height at which the spiral starts, $v_{sw}$ is the velocity of the solar wind, $\Omega_\odot$ is the angular velocity of the Sun, $\phi$ is the longitude, and $\phi_0$ is the longitude at the base of the spiral.

\begin{figure}
   \includegraphics[trim={0cm 0cm 0cm 0cm},clip, width=10cm]{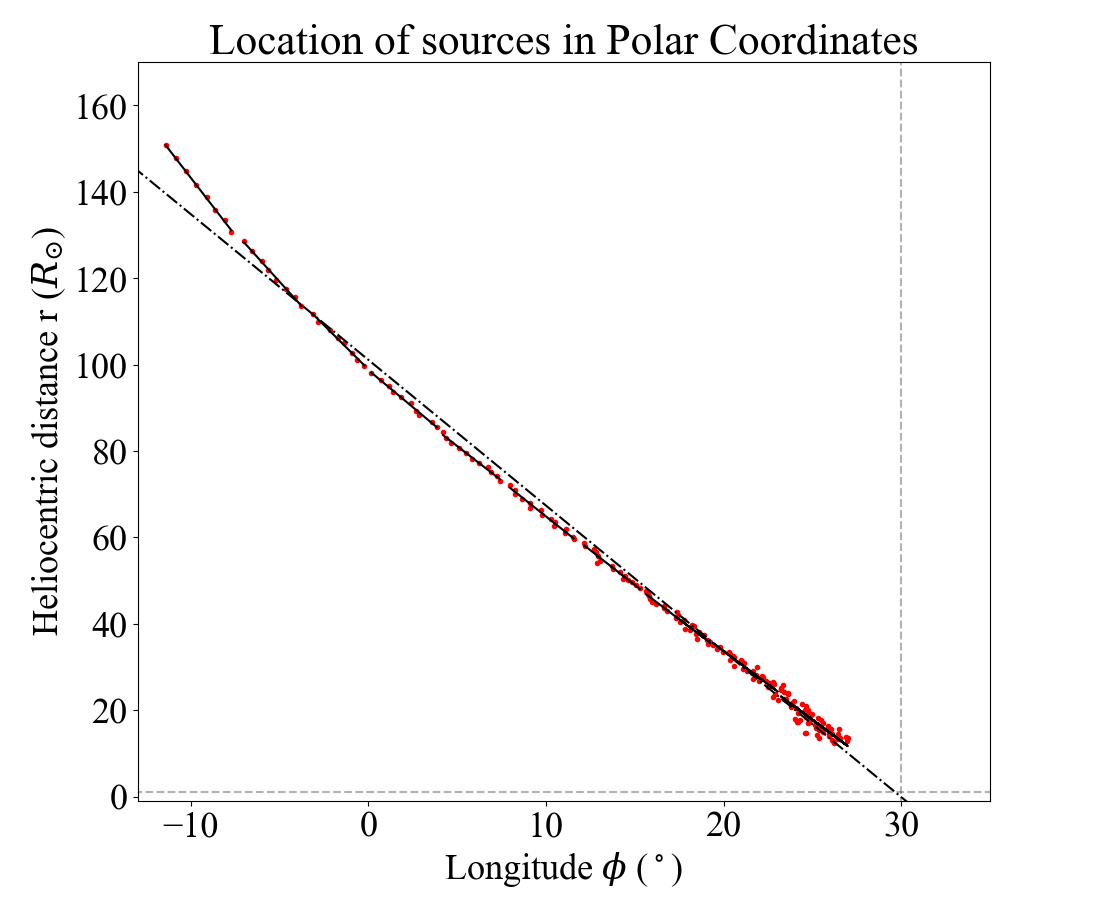}
     \caption{Localisation of the Type III radio burst in polar coordinates. An ideal Parker spiral is shown as a linear function in polar coordinates where the slope is proportional to the speed of the solar wind. A linear piecewise fit was performed in order to obtain the evolution of the wind speed along the source of the emission.}
     \label{fig:resultsRTHETA}
\end{figure}

Equation~\ref{eqn:spiral} shows that an Archimedean spiral is linear in polar coordinates. Using the slope and intercept of the linearised sources in polar coordinates shown in Fig.~\ref{fig:resultsRTHETA}, we derived estimates of the solar wind speed and longitude as $v_{sw}=400 \pm 20$~km~s$^{-1}$ and $\phi_0=30\degr \pm 2\degr$. This implies that there was only a discrepancy of $\sim$~$10\degr$ between the two methods. Comparing these results with Fig.~\ref{fig:solaractivity}, we observed a number of active regions in the $20\degr $ to $ 30\degr$ longitude, suggesting that these ARs are the source of the emission. Despite this uncertainty, both methods still suggest the coronal hole and active region boundary as the likely Type III source region. Closed active region loops may have access to neighbouring open field lines in order to give injected electron beams access to the solar wind. Figure~\ref{fig:resultsBELLA}~(c) also shows the results of performing the Bayesian multilateration of the Type III SRB data obtained using the GH algorithm but without the time evolution fit (Eq.~\ref{eq:timeevo}). The difference between Figs.~\ref{fig:resultsBELLA}~(b) and (c) show that the fitting of Eq.~\ref{eq:timeevo} as a time evolution function to the Type III SRBs is responsible for the smooth Parker spiral seen in Fig.~\ref{fig:resultsBELLA}. This is consistent with the literature that shows that Type III exciters follow a Parker spiral \citep{tri:reiner1998winduly,tri:reiner2009multipoint}, and therefore Eq.~\ref{eq:timeevo} is presented as a characterisation of the leading edge of a Type III SRB.\\

\begin{figure}
\centering 
    \includegraphics[width=8cm]{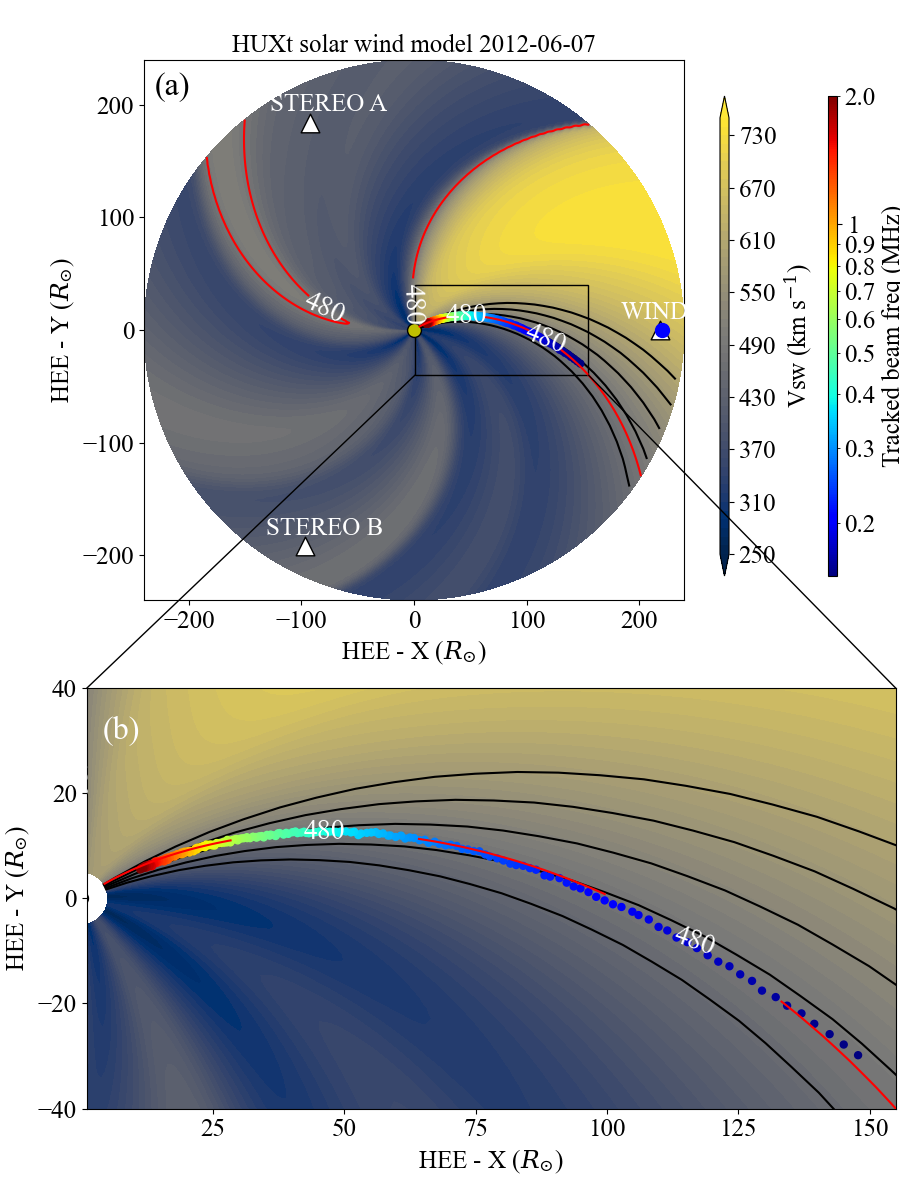}
     \caption{Results from the HUXt solar wind model compared with the sources localised by BELLA. (a) Top down ecliptic view of the multilaterated burst and HUXt results. (b) Zoom-in of the region of interest. The HUXt model shows a region of fast solar wind ($v_{sw}>700$~km~s$^{-1}$) on the +/+ quadrant attributed to the coronal hole CH1 (see Fig.~\ref{fig:solaractivity}~(a)). East of this high speed region is the location of the multilaterated SRB. We found that the SRB sources follow the 480~km~s$^{-1}$ field line with precision ($<5~R_{\odot}$) despite the fact that the HUXt model and the BELLA multilateration are independent of each other. The location of the sources with respect to the region of high solar wind speed is consistent with the location of a cluster of active regions seen in Fig.~\ref{fig:solaractivity}~(b).}
     \label{fig:resultsHUXT}

\end{figure}

To further verify the plausibility of the inferred trajectory in the context of a realistic solar wind, Fig.~\ref{fig:resultsHUXT} shows the BELLA multilateration results overlaid on a solar wind model from the Heliospheric Upwind Extrapolation with time dependence \citep[HUXt;][]{other:owens2020computationally, other:barnard2022huxt}. This model solution was obtained by assimilating the in situ solar wind speeds observed by ACE, STEREO A, and STEREO B to provide a data-constrained solution to the solar wind structure \citep{lang_improving_2021, lang_variational_2019}. This data assimilation was performed over a 27-day window centred on the time of the Type III burst.

The Parker spiral obtained from the BELLA multilateration showed agreement with the HUXt output, suggesting that the electron beam of the emission follows a Parker spiral with a solar wind velocity of 480~km~s$^{-1}$. The HUXt output shows a region of fast solar wind caused by the coronal hole CH1 shown in Fig.~\ref{fig:solaractivity}. This is consistent with the location of the cluster of active regions in Fig.~\ref{fig:solaractivity}~(b), which is adjacent to the eastern region of the CH1 coronal hole region. This further suggests the interface between these active regions and the coronal hole are the origin of the electron beam.

\begin{figure}
   \includegraphics[width=9cm]{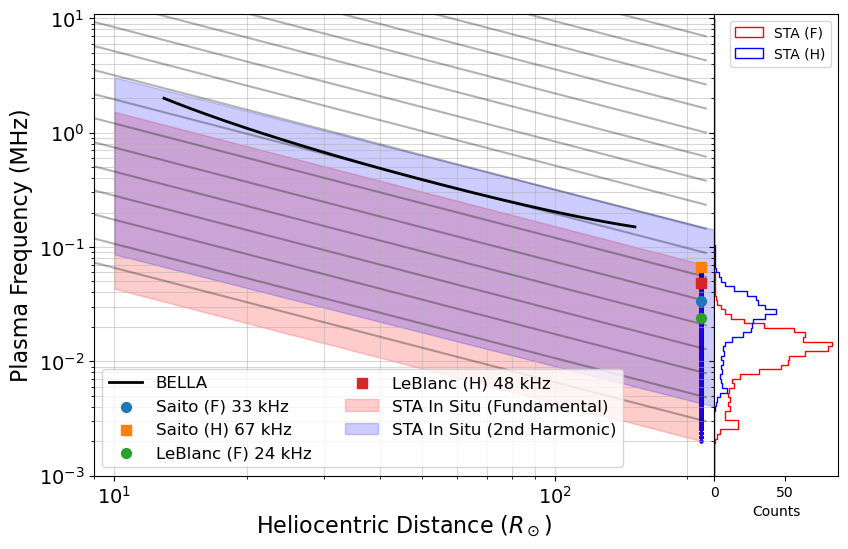}
     \caption{Plasma frequency as a function of heliocentric distance observed by the BELLA multilateration of the Type III SRB and compared against a $1/R^2$ projection of STEREO A in situ electron density data (presented as histograms). Also shown are the harmonic and fundamental values given by the Saito and LeBlanc density models at 1~au. BELLA electron densities are higher than the expected electron densities obtained from models. The highest outliers of the harmonic emission derived from STEREO A in situ data are consistent with BELLA multilaterated sources. An alternative reason for these higher than expected electron densities is scattering \citep{tri:chen2023source, scat:kontar2023anisotropic}.}
     \label{fig:resultsDensities}
\end{figure}

Lastly, Fig.~\ref{fig:resultsDensities} shows the results from plotting the radial distance from the Sun obtained from the BELLA posterior distributions with respect to the frequency of the emission. These frequencies are compared with the plasma frequencies obtained from density models \cite{dm:saito1977} and \cite{dm:Leblanc1998} at 1~au and a $1/R^2$ fundamental and harmonic projection of STEREO A in situ electron density data. We observed that the slope of the BELLA electron density follows an expected $1/R^2$ trend. However, the densities obtained by BELLA compared to the different density models were seen to be significantly larger than expected for either the fundamental or harmonic emission extrapolated inwards from the 1-au density measurements and from the typical density models. With the STEREO A in situ data, which were collected over a Carrington rotation period (28 days) and centred on the time of the burst,  we could state that the observed source locations are consistent with harmonic radiation along a path at the upper end of the distribution of 1-au measured densities for this interval, but this would require the trajectory to lie along a `much more dense than usual' region, which is statistically unlikely given that studies such as \cite{Steinberg1984} have suggested the opposite. This is, however, consistent with \cite{tri:chen2023source} and \cite{scat:kontar2023anisotropic}, which state that radio wave scattering has an effect on the apparent position of SRB exciters, making them appear to be further than their true position. This would imply that the distribution of the speed of propagation $v$ converging at $v=c$ is symptomatic of Type III SRB multilateration being susceptible to converging on the apparent location of the source as lensed by scattering effects. This effect will be the subject of further investigation.

\section{Conclusions}\label{sec:conclusions}
We have presented BELLA, a novel method of positioning Type III radio burst exciters that uses Bayesian statistics as a gateway to perform a detailed error analysis of the uncertainties associated with traditional localisation methods.  
In order to validate BELLA, a simulation was set up under the controlled conditions of a known set of source positions and known propagation speeds. BELLA later recovered the ground truth positions to a better or worse extent depending on spacecraft configuration. We also validated BELLA against a real event, and therefore a Type III radio burst detected by Wind and the STEREO pair was analysed and compared with three other methods of positioning (TDOA \cite{tri:badman2022tracking}; GP \cite{tri:krupar2012goniopolarimetric}; and SEMP \cite{tri:zhang2019forward}) as well as with the HUXt solar wind model \citep{other:owens2020computationally,other:barnard2022huxt}, which we constrained by assimilating the available in situ observations. In order to prepare the data for the Bayesian multilateration, we fitted a GH polynomial to each of the light curves using $1\sigma$ as the rise time detection trigger. These rise time positions were then used to obtain a time evolution function that allowed us to perform the BELLA multilateration despite instrumental discrepancies in cadence, frequency, 
 and resolution as well as differences in frequency channels between Wind/Waves and STEREO/Waves. Despite these instrumental differences, the three spacecraft used in this study have relatively similar spectrogram and orbital specifications. We focused on the highly symmetric case of the STEREO A/B and Wind constellation. Future work will include a less idealised constellation placement and the use of PSP and Solar Orbiter \citep[SolO;][]{sc:marsch2005solar,sc:muller2020solo} data as well as using more than three receivers to obtain full 3D information.\\

In summary, the results obtained from this study are the following:
\begin{enumerate}
    \item Simulations showed that BELLA successfully recovers ground truth data when the spacecraft are evenly separated. We showed that these simulations are capable of distinguishing between regions where the innate multilateration uncertainties are dominant and regions where these innate uncertainties are under the expected instrumental uncertainties.
    \item We localised a Type III radio burst exciter. Results from the BELLA  posterior distributions were found to be in agreement with the output of the analytical TDOA multilateration. BELLA currently uses TOA positioning, and we showed that, in our case, TOA converged on the same apparent source location that TDOA retrieves. This agrees with \cite{tri:kaune2012accuracy}, suggesting that there is no difference in performance between TOA and TDOA. BELLA was also shown to be in agreement with STEREO L3 GP inversions, showing a clear trend between the two and with SEMP, as the localisation of the SEMP method only showed a difference of $\sim$~$10\degr$ from the BELLA Parker spiral. 
    \item  BELLA does not make the $v=c$ assumption a priori, but we found that the MCMC sampling converged on $v\approx c$. This result implies one of two possible outcomes: a)~The effects of refraction and scattering did not play a major role in this event, or b)~only the apparent location of the burst source can be multilaterated. \cite{tri:chen2023source} showed that the latter is likely the case, and scattering will therefore be the subject of further investigation. 
    \item The morphology of the Parker spiral was analysed for the purpose of obtaining an evolution of the solar wind speeds. We found that the multilaterated positions follow an Archimedean Parker spiral of $v_{sw}=400 \pm 20$~km~s$^{-1}$ and $\phi_0=30\degr \pm 2\degr$, which is consistent with the locations of a Coronal hole and the active regions observed on the same date of the event. These source regions were at low latitudes, therefore meaning our 2D treatment of the source location was appropriate.
    \item Output from the HUXt wind model was also compared to the result obtained by BELLA, showing that the BELLA points put the bursts in a solar wind speed of around 480~km~s$^{-1}$. However, the curvature of the track and the model Parker spiral suggest this may have been slightly higher than the true solar wind speed at that location. This $v_{sw}$ is 80~km~s$^{-1}$ higher than the results obtained by fitting an Archimedean Parker spiral to the points obtained by BELLA. 
    \item Density models were compared with the plasma density obtained from the frequencies of the Type III SRB and the radial distance from the Sun. We found that the slope of the linearised density model was in agreement with density models from the literature, following a roughly $1/R^2$ heliospheric trend. The BELLA sources were found to be located at a significantly higher altitude than implied by density models or in situ density measurements at 1~au. This could be explained either by propagation along an anomalously high density filament and harmonic emission or by the localised source being an `apparent' source shifted outwards in radius via scattering \citep{tri:chen2023source, scat:kontar2023anisotropic}. Assuming the latter conclusion, it is interesting to consider the implication that the trajectory still follows a feasible Parker spiral and $1/R^2$ density trend. This may suggest that the lensing effect of scattering may produce a self-similar transformation on the source trajectory and would suggest the inverse transformation may not require a full ray-tracing simulation \citep{scat:kontar2023anisotropic} and could instead be much simpler.
\end{enumerate}

BELLA is a novel method of multilateration that provides an in-depth analysis of the localisation uncertainties, allowing the user to distinguish between the innate multilateration uncertainties and the instrumental and physical uncertainties. BELLA is available as an open source tool for those that want to use it. It is fully capable of operating with more than three spacecraft in 2D. The 3D capabilities of BELLA still have not been developed, and they will be the subject of further work but are a relatively simple addition to the framework outlined in this paper. This will be of special interest when SolO has reached its high orbital inclination $\theta \sim 30\degr$ \citep{sc:muller2020solo}, resulting in a moderate angular separation between receivers and consequently reducing the uncertainty in the $Z$ coordinate.\\

\begin{acknowledgements}
LAC and the research conducted in this publication were supported by the Irish Research Council under grant number GOIPG/2019/2843. DMW's work at the Dublin Institute for Advanced Studies was funded by European Unions Horizon 2020 research and innovation programme under Grant agreement No. 952439 and project number AO 2-1927/22/NL/GLC/ov as part of the ESA OSIP Nanosats for Spaceweather Campaign. DMW's work at Aalto University was funded from the European Research Council (ERC) under the European Unions Horizon 2020 research and innovation programme (project "SYCOS", grant
591 agreement No. 101101005). MJO is part-funded by Science and Technology Facilities Council (STFC) grant number ST/V000497/1. BELLA is available from \url{https://github.com/TCDSolar/BELLA} and \url{https://zenodo.org/records/10276815}. The HUXt solar wind model is available from \url{https://github.com/University-of-Reading-Space-Science/HUXt} and \url{https://zenodo.org/records/7948245}.  The BRaVDA data assimilation scheme from \url{https://github.com/University-of-Reading-Space-Science/BRaVDA} and \url{https://zenodo.org/records/7892408}. The TDOA method compared in figure \ref{fig:resultsBELLA} may be reproduced at code located at \url{https://github.com/STBadman/Radio-Public} and \url{https://zenodo.org/records/10107890}.
\end{acknowledgements}

\bibliography{BELLA.bib}{}
\bibliographystyle{aasjournal}


    
\end{document}